\documentclass[11pt,a4paper]{JHEP3_mod2}

\usepackage{amsmath,latexsym,amssymb,slashed}
\usepackage{graphicx}				
\usepackage{amssymb}					
\usepackage{tabularx}				
\usepackage[vcentermath]{youngtab}	

\newcommand{\eq}[1]{\begin{equation}#1\end{equation}}
\newcommand{\spl}[1]{\begin{split}#1\end{split}}

\newcommand{\boxedeq}[1]{
\begin{equation}
\fbox{
\rule[0.7cm]{0pt}{0pt}
$#1$
\rule[-0.45cm]{0pt}{0pt}
}
\end{equation}
}

\def\d{\text{d}}

\author{Bertrand Sou\`{e}res and Dimitrios Tsimpis\\
Universit\'{e} Claude Bernard (Lyon 1)\\
UMR 5822, CNRS/IN2P3, Institut de Physique Nucl\'{e}aire de Lyon\\
4 rue Enrico Fermi,
F-69622 Villeurbanne Cedex,  France\\

E-mail:
\email{soueres@ipnl.in2p3.fr}, \email{tsimpis@ipnl.in2p3.fr}}
\abstract{
We develop computational tools for calculating supersymmetric higher-order derivative corrections to eleven-dimensional supergravity using the action principle approach.  
We show that, provided the superspace Bianchi identities admit a perturbative solution in the derivative expansion, there are at least two independent superinvariants at the eight-derivative order of eleven-dimensional supergravity. Assuming the twelve-superforms associated to certain anomalous Chern-Simons terms are Weil-trivial, there will be a third independent superinvariant at this order.  Under 
certain conditions, at least two superinvariants will survive to all orders in the derivative expansion. 
However only one of them will be present in the quantum theory: the supersymmetrization of the Chern-Simons terms of eleven-dimensional supergravity 
required for the cancellation of the M5-brane gravitational anomaly by inflow. This superinvariant can be shown to be unique at the eight-derivative order, assuming it is quartic in the fields. 
On the other hand, a necessary  condition for the superinvariant  to be quartic is the exactness, in $\tau$-cohomology, of $X_{0,8}$ --the purely spinorial component of 
the eight-superform related by descent to the M5-brane anomaly polynomial. 
In that case it can also be shown that the solution of the Weil-triviality condition of the corresponding twelve-form, which is a prerequisite for the explicit construction of the superinvariant, is guaranteed to exist. 
We prove that certain highly non-trivial necessary conditions for the $\tau$-exactness of $X_{0,8}$ are satisfied. 
Moreover any potential superinvariant associated to anomalous Chern-Simons terms at the eight-derivative order 
must necessarily contain terms cubic or lower in the fields. 
}
\title{The action principle and the supersymmetrisation of Chern-Simons terms in eleven-dimensional supergravity.}
\preprint{}

\newcommand{\swed}{{\scriptscriptstyle \wedge}}

\begin{document}
\setlength{\parindent}{0pt}



\section{Introduction}

Eleven-dimensional supergravity \cite{Cremmer:1978} is believed to be the low-energy limit of M-theory \cite{Witten:1995:2}, the conjectured 
nonperturbative completion of string theory. 
As such it is expected to receive an infinite tower of higher-order corrections in an expansion in the Planck length or, equivalently, in the derivative expansion. At present such higher-order corrections cannot be systematically constructed within M-theory, so one must resort to indirect approaches. 

One such approach is to calculate the higher-order corrections within perturbative string theory, in particular type IIA in ten dimensions, which is related to eleven-dimensional supergravity by dimensional reduction. 
The effective action of string theory can be systematically constructed perturbatively in a loop expansion in the string coupling,  
\eq{
S_{\mathrm{eff}}=\sum_{g=0}^{\infty}g_s^{2g-2}
\int\d^{10}x\sqrt{G} 
\mathcal{L}_{g}
~, 
}
where $g$ is the loop order (equivalently, the genus of the Riemann surface), $g_s$ is the string coupling constant, $G$ is the spacetime metric  and $\mathcal{L}_{g}$ is the effective action at order $g$. Each $\mathcal{L}_{g}$ admits a perturbative expansion in an infinite series of higher-order derivative terms. Moreover 
it is expected that each $\mathcal{L}_{g}$ should correspond to an independent superinvariant in ten dimensions, see e.g.~\cite{Peeters:2000}.  

The bosonic part of the tree-level effective action takes schematically the following form, 
\eq{\label{12}
\mathcal{L}_{0}
=\mathcal{L}_{\mathrm{IIA}}+\alpha^{\prime3} \big(I_0({R})-\frac18I_1({R}) +\cdots\big)+\mathcal{O}(\alpha^{\prime4})
~,}
where 
$\mathcal{L}_{\mathrm{IIA}}$ is the (two-derivative) Lagrangian of ten-dimensional IIA supergravity, and 
the ellipses stand for terms which have not been completely determined yet.  Unlike 
the case of  $\mathcal{N}=1$ superstrings, the first higher-derivative correction 
starts at order $\alpha^{\prime3}$ (eight derivatives). 
The $I_0$, $I_1$  in (\ref{12}) are defined 
as follows,
\eq{\spl{\label{idef}
I_0(R)&=t_{8}t_{8}{R}^4+\frac12 \varepsilon_{10}t_{8}B{R}^4\\
I_1(R)&=-\varepsilon_{10}\varepsilon_{10}{R}^4+4\varepsilon_{10}t_{8}B{R}^4
~.}}
These were constructed in \cite{deRoo:1993}, to which we refer  for further details, by directly checking 
invariance under part of the supersymmetry transformations. The terms in (\ref{idef}) linear in $B$ are, up to a numerical coefficient, 
Hodge-dual to the Chern-Simons term $B\wedge X_8$ \cite{Vafa:1995,Duff:1995wd}. 
The eight-form $X_8$, see 
(\ref{x8p}) below, is related by descent to the M5-brane anomaly polynomial and is a linear combination of   $(\mathrm{tr}R^2)^2$ and $\mathrm{tr}R^4$. 
Note that the Chern-Simons term drops out of (\ref{12}).

The $R^4$ part of the tree-level effective action was determined in \cite{Green:1981xx,Gross:1986iv} via four-graviton scattering amplitudes and in \cite{Grisaru:1986px,Grisaru:1986vi,Freeman:1986zh} from the 
vanishing of the worldsheet beta-function at four loops. 
The NSNS sector of the four-field part of the effective action (common to all superstring theories in ten dimensions) 
was determined in \cite{Gross:1986mw}: it is  captured by the simple replacement $R\rightarrow\hat{R}$, where $\hat{R}$ is a modified Riemann tensor with torsion 
which includes the NSNS three-form and the dilaton.\footnote{Note that  \cite{Gross:1986mw} 
contains an error that has unfortunately caused some confusion in the literature: the expansion of the $t_8t_8R^4$ terms of eq.~(2.11) in that reference indeed has the form  
of the term in the square brackets on the right-hand side of eq.~(2.13) therein. However, if one replaces $R$ by the modified Riemann tensor $\hat{R}$, given in eq.~(2.12) therein, 
eq.~(2.13) no longer gives the correct expansion of $t_8t_8\hat{R}^4$. 
}    
The $\varepsilon_{10}\varepsilon_{10}{R}^4$ term does not contribute to tree-level four-point scattering amplitudes, but gives a nonvanishing contribution 
to the five-graviton scattering amplitude. 
The complete tree-level four-point effective action for type II superstrings was first determined in \cite{Policastro:2006vt} and, in addition to the NSNS sector, 
 consists of terms of the form $(\partial F )^2\hat{R}^2$ and $\partial^4 F^4$,  where $F$ stands for all RR flux.

The superinvariant   $I_0$  can be further decomposed into two separate 
$\mathcal{N}=1$ superinvariants in ten dimensions \cite{deRoo:1993}, $I_0=-6I_{0a}+24I_{0b}$, where, 
\eq{\spl{
I_{0a}&=(t_8+\frac12 \varepsilon_{10}B)(\mathrm{tr}R^2)^2+\cdots\\ 
I_{0b}&=(t_8+\frac12 \varepsilon_{10}B)\mathrm{tr}R^4+\cdots
~,} }
correspond to the supersymmetrization of the $B\wedge (\mathrm{tr}R^2)^2$ and $B\wedge\mathrm{tr}R^4$ Chern-Simons terms respectively. 
As we show in the following, if the uplift of $I_{0a}$, $I_{0b}$  
gives rise to two separate superinvariants in eleven dimensions, they will necessarily have to be cubic or lower in the fields.

The one-loop effective action takes  the following form \cite{Green:1981xx,Green:1981ya},  
\eq{\label{l1}
\mathcal{L}_{1}
=\alpha^{\prime3} \big(I_0({R})+\frac18I_1({R})+\cdots\big)+\mathcal{O}(\alpha^{\prime4})
~.}
In particular we see that in this case the Chern-Simons term does not drop out, cf.~(\ref{idef}). The ellipses above indicate terms which are not completely known, although partial results exist thanks to five- and six-point amplitude computations 
\cite{Peeters:2001ub,Richards:2008jg,Richards:2008sa,Liu:2013dna}. 
Contrary to the tree-level superinvariant $\mathcal{L}_{0}$ 
which is suppressed at strong coupling, the uplift of the one-loop superinvariant $\mathcal{L}_{1}$ is expected to survive in eleven dimensions, and thus to be promoted to 
an eleven-dimensional superinvariant. We will refer to the latter as {\it the supersymmetrization of the Chern-Simons term} $C\wedge X_8$, the uplift of the ten-dimensional Chern-Simons term, 
where $C$ is the three-form potential of eleven-dimensional supergravity.

An argument of \cite{Howe:2003cy}, which we review in the following, guarantees that if the supersymmetrization of the Chern-Simons term  is quartic or higher in the fields, then it is unique at the eight-derivative order\footnote{The existence of independent superinvariants starting at order higher than eight in the derivative expansion will of course spoil the uniqueness of the 
superinvariant at higher orders.}.  The uniqueness of 
this superinvariant is also supported by the results of \cite{Hyakutake:2005rb,Hyakutake:2006aq,Hyakutake:2007sm} which uses the Noether 
procedure to implement part of the supersymmetry transformations of eleven-dimensional supergravity. The results of these references constrain  the supersymmetrization of the Chern-Simons term to be of the form,
\eq{\label{css}
\Delta \mathcal{L}
=l^{6} \big(t_{8}t_{8}{R}^4-\frac{1}{4!} \varepsilon_{11}\varepsilon_{11}{R}^4 -\frac{1}{6} \varepsilon_{11}t_{8}C{R}^4 + R^3G^2
+\cdots\big)+\mathcal{O}(l^{7})
~,}
where $l$ is the Planck length.  The ellipses indicate terms which were not determined by the analysis of 
\cite{Hyakutake:2005rb,Hyakutake:2006aq,Hyakutake:2007sm}, while the $R^3G^2$ terms were  only partially determined. The reduction of the above 
to ten dimensions is consistent, as expected, with the one-loop IIA superinvariant (\ref{l1}). 
In addition the quartic interactions  $R^2(\partial G)^2$  and $(\partial G)^4$  were determined in \cite{Peeters:2005tb} by eleven-dimensional 
superparticle one-loop computations in the light cone,  
and in \cite{Deser:1998jz,Deser:2000xz,Deser:2005kb} 
by a different method which uses tree amplitudes instead.\footnote{There is disagreement between  \cite{Peeters:2005tb} and \cite{Deser:2000xz}  concerning part of the $(\partial G)^4$ terms.} The $t_{8}t_{8}{R}^4$ terms have also been obtained by four-graviton one-loop amplitudes in eleven dimensions \cite{Green:1997,Russo:1997}, while it can be shown  \cite{Green:1999} that higher loops do not contribute to the superinvariant (\ref{css}).

In the present paper we reexamine the problem of calculating supersymmetric higher-order derivative corrections to eleven-dimensional supergravity from the point of view of the action principle approach. This method relies on the superspace formulation of the theory and is particularly well suited to the supersymmetrization of Chern-Simons terms. 
Given an eleven-dimensional Chern-Simons term  there is an associated gauge-invariant twelve-superform obtained by exterior differentiation. 
The action principle approach can be carried out provided the twelve-form is Weil-trivial, i.e. exact on the space of on-shell  superfields. 
Computing the superinvariant then boils down to explicitly solving the Weil-triviality condition for the twelve-form.

We show that, provided the superspace Bianchi identities admit a perturbative solution in the derivative expansion, there will be at least two independent superinvariants at the eight-derivative order. If we also assume that the twelve-superforms associated to the anomalous (in the presence of an M5-brane) Chern-Simons terms, $C\wedge  (\mathrm{Tr}R^2)^2$ and $C\wedge \mathrm{Tr}R^4$, are separately Weil-trivial, there will be a third independent superinvariant at this order. Moreover we argue that, under certain conditions, at least two of the superinvariants should be expected to survive to all orders in the derivative expansion. 
However only one of those would correspond to the supersymmetrization of $C\wedge X_8$, cf.~(\ref{css}). 

As already noted this superinvariant can be shown to be unique, assuming it is quartic in the fields. On the other hand, a necessary condition for the superinvariant to be quartic is the exactness, in the so-called $\tau$-cohomology, of $X_{0,8}$ --the purely spinorial component of $X_8$. In that case we also show that the solution of the Weil-triviality condition of the corresponding twelve-form is guaranteed to exist. 
Proving the $\tau$-exactness of $X_{0,8}$ is the first and arguably most difficult step in obtaining the explicit solution to the Weil-triviality condition of the twelveform, and 
therefore constructing the superinvariant using the action principle. 

To tackle this computationally intensive problem we have built on the computer program \cite{Gran:2001:2}, 
to supplement it, among other things, with functionalities related to Young tableaux \cite{Bertrand}. 
By a combination 
of calculational techniques involving the implementation 
of Fierz identities and Young tableaux projections  we   
prove that certain highly non-trivial necessary conditions for the $\tau$-exactness of $X_{0,8}$ are satisfied. As a corollary of our work, it follows that any potential superinvariant associated to the anomalous Chern-Simons terms, $C\wedge  (\mathrm{Tr}R^2)^2$ and $C\wedge \mathrm{Tr}R^4$, must necessarily contain terms cubic or lower in the fields.

The plan of the rest of the paper is a follows. In section \ref{sec:cohomology} we review the different superspace cohomologies that will be useful in the following. In section \ref{sec:ap} 
we introduce the action principle approach and  in section \ref{sec:cjs} we  show how to obtain the eleven-dimensional supergravity of \cite{Cremmer:1978} in this framework. In section \ref{sec:five} we apply the action principle to derive the five-derivative correction. Section \ref{sec:eight} considers the eight-derivative correction. In section \ref{sec:invariants} we examine the number of independent superinvariants at the eight-derivative order. Section \ref{sec:x8} addresses the problem of the $\tau$-exactness of $X_{0,8}$. In section \ref{sec:integrability} we discuss the conditions for the existence of the superinvariants to all orders in the perturbative expansion. We conclude in section \ref{sec:disc}. Further technical details are  included in the appendices.

\section{Cohomology in superspace}\label{sec:cohomology}

In this section we review the various superspace cohomology groups that will be useful in the following. 
This is not new material, but we are including it here to make the paper self-contained and for the benefit of   
the readers who may not be familiar with the relevant literature.

Let us start by explaining our conventions: 
Eleven-dimensional superspace \cite{Cremmer:1980,Brink:1980} consists of eleven even (bosonic) and thirty-two odd
(fermionic) dimensions, with structure group the
eleven-dimensional spin group. 
Let $A=(a,\alpha)$ be flat tangent superindices, where $a=0,\ldots 10$ is a Lorentz vector 
index and $\alpha=1,\ldots 32$ is a Majorana spinor index. 
Curved superindices will be denoted by 
$M=(m,\mu)$, with the corresponding supercoordinates denoted by 
$Z^M=(x^m,\theta^\mu)$. The supercoframe  is denoted by 
$E^A=(E^a,E^{\alpha})$  while the dual superframe is denoted by
$E_A=(E_a, E_{\alpha})$. 
We can pass from the coframe to the
coordinate basis using the supervielbein, $E^A=\d z^M E_M{}^A$.

We shall assume the existence of a connection one-form $\Omega_A{}^B$ with values in the Lie algebra of
the Lorentz group. 
In particular this implies that,
\eq{\label{lore}
\Omega_{(a}{}^c\eta_{b)c}=0~, ~~~
\Omega_\alpha{}^\beta={1\over4}(\gamma^a{}_b)_\alpha{}^\beta\Omega_a{}^b~,
~~~\Omega_a{}^\beta=0=\Omega_\alpha{}^b~. 
}
The associated supertorsion and supercurvature tensors are then given by:
\eq{\spl{
T^A &=DE^A:=\d E^A + E^B \swed\Omega_B{}^A=\frac12 E^C \swed E^B T_{BC}{}^A \\
R_A{}^B &=\d\Omega_A{}^B +\Omega_A{}^C\swed\Omega_C{}^B=\frac12 E^D \swed E^C R_{CDA}{}^B~,
\label{2.1}
 }}
where the exterior derivative is given by $\d=\d z^M\partial_M$. 
The assumption of a Lorentzian structure group implies that the components of the  curvature two-form obey 
a set of equations analogous to (\ref{lore}). 
The super-Bianchi identities (BI) for the torsion and the curvature, 
\eq{\spl{
DT^A&=E^B\swed R_B{}^A~,\\ 
DR_A{}^B&=0~,
\label{SSBI}
}}
follow from the definitions (\ref{2.1}). Moreover, a theorem due to Dragon \cite{Dragon:1979nf} ensures  
that for a Lorentz structure group the second BI above follows from the first and need not be considered separately. 
Once constraints are imposed the BI cease to be automatically satisfied. As was shown in \cite{Brink:1980}, by imposing the conventional 
constraint 
\eq{\label{conv}T^c_{\alpha\beta} =i\gamma^c_{\alpha\beta} ~,}
and solving the torsion BI,  
one recovers ordinary eleven-dimensional supergravity. In particular one determines in this way all components of the torsion. 
In addition one can construct a closed super four-form $G_4$ and a super seven-form $G_7$ obeying \cite{DAuria:1982uck,Candiello:1994},\footnote{The $G_7$ BI receives a correction at the eight-derivative order, cf. (\ref{gmod}) below.}
\eq{\label{gbi}  
\d G_4=0~,~~~
\d G_7+\frac12 G_4\swed G_4=0
~,}
whose bosonic components correspond to the 
eleven-dimensional supergravity four-form and its Hodge-dual, respectively:
\eq{G_{m_1\ldots m_7}=(\star G)_{m_1\ldots m_7}~.}
The solution of the eleven-dimensional superspace BI is reviewed in appendix \ref{app:ss}.

\subsection{De Rham cohomology and Weil triviality}\label{sec:dr}

Let $\Omega^n$ be the space of $n$-superforms. 
Thanks to the nilpotency of the exterior superderivative, one can define de Rham cohomology groups in superspace in the same way as in the case of bosonic space,
 \eq{
 H^{n}=\{\omega\in \Omega^{n}|\d\omega=0\} / \{ \omega\sim\omega+\d\lambda,
 \lambda\in \Omega^{n-1}\}~.
 }
The fact that the topology of the odd directions is trivial means that the de Rham cohomology of a supermanifold coincides with 
the de Rham cohomology of its underlying bosonic manifold, also known as the {\em body} of the supermanifold. 
In the remainder of the paper we shall assume that the body has trivial topology. This is the simplest type of supermanifold, sometimes called a 
{\em graded manifold}. It implies in particular that every $\d$-closed superform 
is $\d$-exact. 

There is an important caveat to the previous statement: it is only valid when the cohomology is computed on the space of unconstrained 
superfields. Once constraints are imposed it ceases to be automatically satisfied. 
Adopting the terminology of \cite{Bonora:1986xd}, we shall call {\em Weil-trivial} those $\d$-closed superforms which are also $\d$-exact on the space of constrained (also referred to as ``on-shell'', or ``physical'') superfields. The cohomology 
groups computed on the space of constrained superfields will be denoted by $H^{n}(\rm{phys})$, as in \cite{Howe:2003cy}. As already emphasized, there is no a priori reason 
why $H^{n}(\rm{phys})$ should coincide with the cohomology of the  body of the supermanifold.

\subsection{$\tau$-cohomology}\label{sec:tc}

The space of superforms can be further graded according to the even, odd 
degrees of the forms. We denote the space of
forms with $p$ even and $q$ odd components by
$\Omega^{p,q}$ 
so that,
\eq{\Omega^n=\oplus\!\!\!\!\sum_{p+q=n}\!\!\Omega^{p,q} ~.}
 A $(p,q)$-superform $\omega\in\Omega^{p,q}$ can be expanded as follows,
 \eq{
 \omega=\frac{1}{ p! q!}E^{\beta_q}\ldots E^{\beta_1}\, E^{a_p}\ldots E^{a_1} \omega_{a_1\ldots
 a_p \beta_1\ldots \beta_q}~.
 \label{3.2}
 }
In the following we will use the notation $\Phi_{(p,q)}\in\Omega^{p,q}$ for the projection of a superform 
 $\Phi\in\Omega^{n}$ onto its $(p,q)$ component.

The exterior superderivative, $\d:\Omega^{p,q}\rightarrow
\Omega^{p+1,q}+\Omega^{p,q+1} +\Omega^{p-1,q+2}+\Omega^{p+2,q-1}$, when written out in this basis 
will give rise to components of the torsion as it acts on the coframe. Following
\cite{Bonora:1990mt} we split $\d$ into its various components with
respect to the bigrading,
 \eq{
\d=\d_b + \d_f + \tau + t~,
 \label{3.3}
 }
where $\d_b$, $\d_f$ are even, odd derivatives respectively, such that 
$\d_b:\Omega^{p,q}\rightarrow\Omega^{p+1,q}$, $\d_f:\Omega^{p,q}\rightarrow\Omega^{p,q+1}$. The operators 
 $\tau$ and $t$ are purely algebraic and can be expressed in terms of the torsion.  
Explicitly, for any $\omega\in \Omega^{p,q}$ we have,
\eq{\spl{
(\d_b \omega)_{a_1 \dots a_{p+1} \beta_1 \dots \beta_q} &= (p+1) \Big( {D}_{[a_1} \omega_{a_2\dots a_{p+1}] \beta_1 \dots \beta_q} + {p\over2} {T_{[a_1 a_2|}}^c \; \omega_{c| a_3 \dots a_{p+1}] \beta_1 \dots \beta_q} \\
&\phantom{=} + q (-1)^p \; {T_{[a_1 | (\beta_1|}}^{\gamma} \; \omega_{|a_2 \dots a_{p+1}] \gamma|\beta_2 \dots \beta_q)} \Big) \\
\\
(\d_f \omega)_{a_1 \ldots a_{p} \beta_1 \ldots \beta_{q+1}} &= (q+1) \Big( (-1)^p {D}_{(\beta_1|} \omega_{a_1 \ldots a_{p} |\beta_2 \ldots \beta_{q+1})} + {q\over2} {T_{(\beta_1 \beta_2|}}^{\gamma} \; \omega_{a_1 \ldots a_{p} \gamma|\beta_3 \ldots \beta_{q+1})} \\
&\phantom{=} + p(-1)^p \; T_{(\beta_1| [a_1|}{}^{c}\omega_{c|a_2 \ldots a_{p}] |\beta_2 \ldots \beta_{q+1})} \Big) \\
\\
(\tau\omega)_{a_1 \ldots a_{p-1} \beta_1 \ldots \beta_{q+2}} &= {1\over2} (q+1)(q+2) \; {T_{(\beta_1\beta_2|}}^c \; \omega_{c a_1 \ldots a_{p-1} |\beta_3 \ldots \beta_{q+2})} \\
\\
(t\omega)_{a_1\ldots a_{p+2}\beta_1\ldots \beta_{q-1}} &= {1\over2} (p+1)(p+2) \; {T_{[a_1 a_2}}^{\gamma} \; \omega_{a_3 \ldots a_{p+2}] \gamma \beta_1 \ldots \beta_{q-1}}~.
\label{derivs}
}}
The nilpotency of the exterior derivative, $\d^2=0$, implies the following identities:
 \eq{\spl{\label{3.4}
 \tau^2 &= 0\\
 \d_f\tau+ \tau \d_f&=0\\
 \d_f^2 + \d_b\tau + \tau \d_b&=0\\
 \d_b \d_f + \d_f \d_b + \tau t+ t\tau&=0\\
 \d_b^2 +\d_ft + t \d_f &=0\\
 \d_bt + t \d_b &=0\\
 t^2 &=0~.
 }}
The first and the last of these equations are algebraic identities and are 
always satisfied. On the other hand, as a consequence of the splitting of the tangent bundle into 
even and odd directions, the remaining identities 
are only satisfied provided the torsion
tensor obeys its Bianchi identity.

The first of the equations in (\ref{3.4}), the nilpotency of the $\tau$ operator, implies that we can consider the cohomology of
$\tau$, as first noted in \cite{Bonora:1990mt} (see also \cite{DAuria:1982uck} for some related concepts). Explicitly we set,
 \eq{
 H^{p,q}_{\tau}=\{\omega\in \Omega^{p,q}|\tau\omega=0\} / \{ \omega\sim\omega+\tau \lambda,
 \lambda\in \Omega^{p+1,q-2}\}~.
 }
As in the case of de Rham cohomology, one could make a distinction between cohomology groups computed on the 
space of unconstrained superfields and those computed on the space of physical fields. 

Suppose now that the conventional constraint (\ref{conv}) is imposed so that $\tau$ reduces to a gamma matrix. 
It was conjectured in \cite{Howe:2003cy}, consistently with the principle of maximal propagation of \cite{Cederwall:2001dx}, 
that in this case the only potentially nontrivial $\tau$-cohomology appears as a result of the so-called M2-brane identity,
\eq{(\gamma^a)_{(\alpha_1\alpha_2} (\gamma_{ab})_{\alpha_3\alpha_4)}=0 \label{5.9} 
~.}
Explicitly, for $p=0,1,2$, one may form the following $\tau$-closed $(p,q)$-superforms,
\eq{\omega_{\alpha_1\ldots \alpha_q}=S_{\alpha_1\ldots \alpha_q} ~;~~~
\omega_{a\alpha_1\ldots \alpha_q}=(\gamma_{ab})_{(\alpha_1\alpha_2}P^b{}_{\alpha_3\ldots \alpha_q)}~;~~~
\omega_{ab\alpha_1\ldots \alpha_q}=(\gamma_{ab})_{(\alpha_1\alpha_2}U_{\alpha_3\ldots \alpha_q)}
~,}
with $S$, $P$, $U$, arbitrary superfields.  It can be seen using (\ref{5.9}) that 
the forms $\omega$ above correspond to nontrivial elements  of $H^{p,q}_{\tau}$ with $p=0,1,2$. The conjecture of  
\cite{Howe:2003cy} means that all nontrivial cohomology is thus generated, and that all  $H^{p,q}_{\tau}$ groups are trivial for $p\geq 3$. 
This was subsequently proven in  \cite{Movshev:2011pr} and \cite{Brandt:2009xv,Brandt:2010fa,Brandt:2010tz,Brandt:2013xpa}.

\subsection{Spinorial cohomology}\label{sec:sc}

Following \cite{Howe:2003cy}, let us  now define a spinorial derivative $\d_s$ which acts on
elements of $\tau$-cohomology, $\d_s:H^{p,q}_{\tau}\rightarrow H^{p,q+1}_{\tau}$. For any $\omega\in[\omega]\in H^{p,q}_{\tau}$ we set,
 \eq{
 \d_s[\omega]:=[\d_f \omega]
 \label{3.11}
 ~.}
To check that this is well-defined, one first shows that $\d_f\omega$ is
$\tau$-closed, 
\eq{\tau\d_f\omega=-\d_f\tau\omega=0
~,}
where we used the second equation in (\ref{3.4}).  
Morever $\d_s [\omega]$ is independent of the choice of
representative,
\eq{[\d_f(\omega+\tau\lambda)]=
[\d_f\omega-\tau\d_f\lambda]
=[\d_f\omega]
~.}
Furthermore it is simple to check that $\d_s^2=0$,
\eq{
\d^2_s[\omega]=\d_s[\d_f \omega]=
[\d^2_f \omega]=-[(\d_b\tau + \tau \d_b)\omega]
=0
~,}
where we took into account the third equation in (\ref{3.4}). 
We can therefore define the corresponding spinorial cohomology groups 
$H^{p,q}_s$ as follows,
 \eq{
 H^{p,q}_{s}=\{\omega\in H_{\tau}^{p,q}|\d_s \omega=0\} / \{\omega\sim\omega+\d_s \lambda,
 \lambda\in H_{\tau}^{p,q-1}\}~.
 \label{scoh}
 }
The notion of spinorial cohomology was originally introduced in \cite{Cederwall:2001bt,Cederwall:2001dx} and applied in a series of papers 
with the aim of computing higher-order corrections to supersymmetric theories \cite{Cederwall:2001td,Cederwall:2002df,Howe:2003sa,Tsimpis:2004rs,Cederwall:2004cg}, and more 
recently in \cite{Movshev:2014hha,Chang:2014kma,Chang:2014nwa}. 
The spinorial cohomology as presented above was introduced in \cite{Howe:2003cy} and is independent of the value of the 
dimension-zero torsion. It reduces to the spinorial cohomology of \cite{Cederwall:2001bt,Cederwall:2001dx} upon imposing the conventional constraint (\ref{conv}).

\subsection{Pure-spinor cohomology}

It was first pointed out by P.~Howe \cite{Howepc} and subsequently elaborated in \cite{Howe:2003cy}, that
in the case where the dimension-zero torsion is flat, cf.~(\ref{conv}), the cohomology groups $H_s^{0,q}$ are
isomorphic to Berkovits's pure-spinor cohomology groups \cite{Berkovits:2002uc}. Therefore, in view of what was said in section \ref{sec:sc},  the latter are also isomorphic to the
spinorial cohomology groups that had been computed a few months earlier in \cite{Cederwall:2001dx}. In the following we briefly explain 
the equivalence between the two formulations.

The pure spinor cohomology groups are defined as follows. Consider an eleven-dimensional pure spinor, $\lambda^\alpha$, \`{a} la Berkovits, i.e. such that it 
obeys,\footnote{This definition is different from an eleven-dimensional pure spinor  \`{a}  la Cartan, used in \cite{Howe:1991}, which obeys $\lambda^\alpha\gamma^{ab}_{\alpha\beta}\lambda^\beta=0$ in addition to (\ref{psb}).}
\eq{\label{psb}\lambda^\alpha\gamma^a_{\alpha\beta}\lambda^\beta=0
~.}
The pure spinor $\lambda^\alpha$ is assigned ghost number one. 
Furthermore we define a form of ghost number $q$ as a multi-pure spinor,
\eq{\label{psb1}\omega=\lambda^{\alpha_1}\ldots\lambda^{\alpha_q}\omega_{\alpha_1\ldots\alpha_q}~.}
Note that the above definition implies that $\omega\in[\omega]\in H^{0,q}_{\tau}$: indeed shifting $\omega_{\alpha_1\ldots\alpha_q}$ by a  $\tau$-exact term would 
drop out of  the right-hand side above due to the contractions with the pure spinors; moreover $\omega_{\alpha_1\ldots\alpha_q}$ is trivially $\tau$-closed.
 
The pure-spinor BRST operator is defined as follows,
\eq{Q:=\lambda^{\alpha}D_{\alpha}~,}
where  $D_{\alpha}$ is the spinor component of the covariant derivative defined in flat superspace.  
Therefore the action of $Q$ on omega, 
\eq{Q\omega=\lambda^{\alpha_1}\ldots\lambda^{\alpha_q}\lambda^{\alpha_{q+1}}D_{\alpha_{q+1}}\omega_{\alpha_1\ldots\alpha_q}~,} 
corresponds precisely to 
the action of $d_s$ defined in (\ref{3.11}). Indeed, for flat superspace the torsion terms drop out and  $d_f$ reduces to $D_\alpha$, cf. the second line of 
eq.~(\ref{derivs}). Moreover the 
contraction with the pure spinors on the right-hand side above implies that $Q\omega\in[Q\omega]\in H^{0,q+1}_{\tau}$, for the same reasons noted below (\ref{psb1}). 
In other words, in the 
linearized limit the pure-spinor cohomology groups of ghost number $q$ are isomorphic to the spinorial cohomology groups $H^{0,q}_{s}$. For an extended review of pure-spinor superfields, see \cite{Cederwall:2013vba}.

\section{The action principle}\label{sec:ap}

The {\em action principle}, also known as {\em ectoplasmic integration},  
is a way of constructing superinvariants in $D$ spacetime dimensions as integrals of 
closed $D$-superforms \cite{DAuria:1982pm,Gates:1997ag}. Indeed if $L$ is a closed
$D$-superform,  the following action is invariant under supersymmetry, 
\eq{
 S=\frac{1}{D!}\int\, \d^Dx\, \varepsilon^{m_1\ldots m_D} L_{m_1\ldots m_D}|~,
 \label{5.6}
 }
where a vertical bar denotes the evaluation of a superfield at $\theta^{\mu}=0$.
This can be seen as follows. Consider an 
infinitesimal super-diffeomorphism generated by a super-vector field $\xi$. 
The corresponding 
transformation of the action reads,
 \eq{
 \delta L = \mathcal{L}_\xi L= (\d i_\xi + i_\xi \d) L=\d i_\xi L~,
 \label{5.7}
 }
where we took into account that $L$ is closed. On the other hand,   
local supersymmetry transformations and spacetime 
diffeomorphisms are generated by $\xi|$ and, in view of (\ref{5.7}), the integrand in (\ref{5.6}) transforms as a
total derivative under such transformations. The action is thus invariant assuming boundary terms can be neglected.

This method is particularly well-suited to actions with 
Chern-Simons (CS) terms and indeed has  been used to 
construct all Green-Schwarz brane actions  \cite{Howe:1998tsa,Bandos:1995dw}, see \cite{Kuzenko:2012ew,Kuzenko:2013rna} for more recent applications to other theories and \cite{Berkovits:2008qw} for applications to higher-order corrections. 
The idea is as follows: let $Z_D$ be the CS term 
and $W_{D+1}=\d Z_D$ its exterior derivative. Obviously  $W_{D+1}$ is a closed form. 
On the other hand one might be led to conclude that the de Rham cohomology group of rank $D+1$ must be trivial on a supermanifold 
whose body is $D$-dimensional, hence  $W_{D+1}$  must also be exact. This means that it can be written 
as $W_{D+1}=\d K_D$ where now, contrary to $Z_D$, $K_D$ is a globally-defined (gauge-invariant) superform. 
It follows that $L_D:=Z_D-K_D$ is a closed superform, and can therefore be used to construct a supersymmetric action 
as in (\ref{5.6}). 

Eleven-dimensional supergravity is another example of an action with Chern-Simons terms, and we turn to the application of the action principle to this 
case in the following sections. Unfortunately there is a caveat to the previous argument that $W_{D+1}$ is exact. As already noted in section 
\ref{sec:dr}, this argument can be applied only in the case where the cohomology is computed on the space of unconstrained superfields, but is not {\em a priori} true on the space of 
physical (on-shell) superfields. Interestingly it does turn out to be true in all known cases. As we will see in the following this includes the case of ordinary 
eleven-dimensional supergravity as well as its supersymmetric corrections with five derivatives.  In section \ref{sec:x8} we show that a sufficient condition for the 
Weil triviality of the eight-derivative correction is the $\tau$-exactness of $X_{0,8}$.

We shall parameterize the derivative expansion in terms of the Planck length $l$, so that the Cremmer-Julia-Scherk two-derivative action (CJS) corresponds to zeroth order in $l$. 
In section \ref{sec:eight} we show that, provided the four- and seven-form BI are satisfied at order $\mathcal{O}(l^6)$, cf.~(\ref{gmod}), there are at least 
two Weil-trivial twelveforms $W_{12}$ and hence at least two independent  supersymmetric actions with eight derivatives. 
Provided the twelve-forms associated to certain anomalous CS terms are Weil-trivial, cf. (\ref{uvd}) below, 
there will be a third independent superinvariant at this order. We argue that 
at least two of those superinvariants will exist to all orders in the derivative expansion.

As we will see in detail in the following, in practice one solves for the flat components of the closed superform $L_{D}$ in a stepwise fashion in increasing  
engineering dimension. Once all flat components of $L_{D}$ have been determined in this way, the explicit form of the action (\ref{5.6}) can be extracted using the formula, 
\eq{\spl{\label{az}
L_{m_1\ldots m_D}\big|=e_{m_D}{}^{a_D}\cdots e_{m_{1}}{}^{a_{1}}L_{a_1\ldots a_D}\big|+D ~\!e_{m_{D}}{}^{a_{D}}\cdots e_{m_2}{}^{a_2}&\psi_{m_1}{}^{{\alpha}_1}L_{{\alpha}_1 a_2\ldots a_D}\big| +\cdots\\
&\cdots+\psi_{m_D}{}^{{\alpha}_D} \cdots  \psi_{m_1}{}^{{\alpha}_1}L_{{\alpha}_1 \ldots \alpha_D}\big|
~,}}
where $\psi_m^\alpha := E_m{}^\alpha |$ and $e_m{}^a := E_m{}^a |$ are identified as the gravitino and the vielbein of (bosonic) spacetime respectively. In particular  the bosonic terms of the Lagrangian can be read off immediately from $L_{a_1\ldots a_D}$.

\subsection{CJS supergravity in the action principle formulation}\label{sec:cjs}

The eleven-dimensional supergravity action reads \cite{Cremmer:1978},
\eq{\label{ns} {S} = \int \; \big( R \star 1 -\frac12 \; G_4 \swed \star G_4 -\frac16 \left.C_3 \swed G_4 \swed G_4\big)\right|~,
}
where $\d C_3=G_4$ is the threeform potential; it is understood that only the bosonic $(11,0)$ components of the forms 
enter the formula above, as in (\ref{5.6}). 

This action can also be understood from the point of view of the action principle as follows. 
The twelveform corresponding to the CS term reads,
\eq{W_{12}= -\frac16 G_4 \swed G_4 \swed G_4 =\d Z_{11}~;~~~
Z_{11}=-\frac16 C_3 \swed G_4 \swed G_4
~.}
Using the BI  (\ref{gbi}) this can also be written in a manifestly Weil-trivial form, 
\eq{\label{ty}W_{12}=\d K_{11} ~;~~~
 K_{11}=\frac13 G_4 \swed G_7 
 ~.}
Taking $L_{11}=Z_{11}-K_{11}$ we  obtain that the following action is invariant under supersymmetry,
\eq{ S = \int \; \big(  -\frac13 G_4 \swed G_7
-\frac16 \left.C_3 \swed G_4 \swed G_4
\big)\right|~.
}
This can then be put in the form (\ref{ns}) by using the on-shell conditions $\star G_4=G_7$ and 
$G_4 \swed \star G_4=6R \star 1$, cf.~appendix  \ref{app:0}. Therein we also give the details of the solution of the superspace equation $W_{12}=\d K_{11}$ and we show, as 
a byproduct,  that the solution for $K_{11}$ given in  (\ref{ty}) is unique up to exact terms.

\subsection{The $\mathcal{O}(l^3)$ correction (five derivatives)}\label{sec:five}

It was shown in \cite{Tsimpis:2004rs}, by directly computing the relevant spinorial cohomology group, that  there is a unique superinvariant at the five derivative level 
(order $l^3$ in the Planck length).\footnote{As explained in \cite{Tsimpis:2004rs}, on a topologically trivial spacetime
manifold  this superinvariant can be
removed by an appropriate field redefinition of the threeform superpotential. However 
on a spacetime with nonvanishing first Pontryagin class the superinvariant cannot be
redefined away without changing the quantization condition of the fourform field strength.} The modified 
eleven-dimensional action to order $l^3$ reads,
\eq{\label{nsf} {S} = \int \; \Big( R \star 1 -\frac12 \; G_4 \swed \star G_4 - \frac16 C_3 \swed G_4 \swed G_4
+l^3
\big( \left.C_3 \swed G_4 \swed  \mathrm{tr}R^2 +2~\!\mathrm{tr}R^2\swed \star G_4 
   \big)\Big)\right\vert
~,}
where an arbitrary numerical coefficient has been absorbed in the definition of $l$ and $\mathrm{tr}R^2:=R_a{}^b\swed R_b{}^a$; 
it is understood that only the bosonic $(11,0)$ components of the forms 
enter the formula above. 
This action can also be easily understood from the point of 
view of the action principle as follows. 
Consider the twelve-form corresponding to the CS term at order $l^3$,
\eq{W_{12}=    G_4 \swed G_4 \swed  \mathrm{tr}R^2 =\d Z_{11}~;~~~
Z_{11}=   C_3 \swed G_4 \swed  \mathrm{tr}R^2
~.}
Using the BI  (\ref{SSBI}), (\ref{gbi}) this can also be written in a manifestly Weil-trivial form, 
\eq{\label{tyf}W_{12}=\d K_{11} ~;~~~
 K_{11}=-2~\!G_7 \swed  \mathrm{tr}R^2 
 ~.}
Taking $L_{11}=Z_{11}-K_{11}$ we  obtain the following superinvariant at order $l^3$,
\eq{ \Delta S = \int \; \big(
 C_3 \swed G_4 \swed  \mathrm{tr}R^2+2\left.G_7 \swed  \mathrm{tr}R^2
\big)\right|~.
}
This can be seen, using the Hodge duality relation $G_7=\star G_4$, to precisely correspond to the order-$l^3$ terms in (\ref{nsf}).\footnote{
The Hodge duality relation between $G_7$ and $G_4$ is expected to receive higher-order corrections (see below \ref{411}). These can be neglected here since $\Delta S$ is already 
a higher-order correction.}

In appendix \ref{app:3} we work out in detail the superspace equation $W_{12}=\d K_{11}$ and confirm 
that the solution (\ref{tyf}) for $K_{11}$ is unique up to exact terms, in accordance with the spinorial cohomology result  
of  \cite{Tsimpis:2004rs}.

\section{The $\mathcal{O}(l^6)$ correction (eight derivatives)}\label{sec:eight}

As was shown in \cite{Duff:1995wd,Freed:1998tg}, the requirement that the M5-brane gravitational anomaly is cancelled by inflow 
from eleven dimensions implies the existence of certain CS terms $Z_{11}$ at the eight-derivative order 
in the eleven-dimensional theory. 
The corresponding twelve-form reads,
 \eq{
 W_{12}=  l^6 G_4 \swed X_8 = \d Z_{11}~;~~~ Z_{11}=l^6  C_3 \swed X_8
 ~,
 \label{5.10}}
where $X_8$ is related to the M5-brane anomaly polynomial  by descent,
\eq{\label{x8p}
X_8=\mathrm{tr}R^4-\frac14 (\mathrm{tr}R^2)^2
~,}
and we have set $(\mathrm{tr}R^2)^2:=\mathrm{tr}R^2\swed \mathrm{tr}R^2$, $\mathrm{tr}R^4:=R_a{}^b\swed R_b{}^c\swed R_c{}^d\swed R_d{}^a$. 
At eight derivatives the modified four- and seven-form BI read,
\eq{\label{gmod}\d G_4=0~;~~~\d G_7+\frac12 G_4 \swed G_4=l^6 X_8~,}
where a numerical coefficient has been absorbed in the definition of $l$. 
We expand the forms perturbatively in $l$,
\eq{\label{pert}
G_4=G_4^{(0)}+ l^6 G_4^{(1)}+\cdots~;~~~
G_7=G_7^{(0)}+ l^6 G_7^{(1)}+\cdots
~,}
and similarly for the supercurvature $R_A{}^B$.  
Note that in the expansion above the bosonic components of the lowest-order fields, $G^{(0)}_{m_1\ldots m_4}$ etc, are identified with the fieldstrengths 
of the supergravity multiplet, while 
the higher-order fields $G_4^{(1)}$ etc, are composite higher-derivative fields which are polynomial in the fieldstrengths of the supergravity fields.

Solving perturbatively the BI at each order in $l$, taking into account that the exterior superderivative 
$\d=\d z^M \partial_M$ is zeroth-order in $l$,  implies,
 \eq{\spl{\label{gmodp}
  \d G_4^{(0)}= 0~&;~~~  \d G_4^{(1)} = 0~;\\
  \d G_7^{(0)} +\frac12 G_4^{(0)} \swed G_4^{(0)} = 0~&;~~  \d G_7^{(1)} + G_4^{(1)}\swed G_4^{(0)} = X_8^{(1)}
~,}}
where we have set $l^6X_8=l^6X_8^{(1)}+\cdots$. Note that $X_8^{(1)}$ only involves the lowest-order curvature $R^{(0)}$. 
Let us expand the twelve-form $W_{12}$ 
perturbatively in $l$, $W_{12}=l^6 W_{12}^{(1)}+\cdots$,  so that,
\eq{\label{48}
W^{(1)}_{12} =  X_8^{(1)}\swed G_4^{(0)} = \d Z_{11}~;~~~ 
Z_{11}=   X_8^{(1)}\swed C_3^{(0)}
~.}
It then follows from (\ref{gmodp}) that this can also be written in a manifestly Weil-trivial form as follows, 
\eq{\label{49}
W^{(1)}_{12} = \d K_{11}~;~~~ K_{11}= G_7^{(1)} \swed G_4^{(0)}- 2G_7^{(0)}\swed G_4^{(1)}
~.}
In particular we see that it suffices to solve the four- and seven-form BI in order to determine the order-$l^6$ 
superinvariant corresponding to $L_{11}=Z_{11}-K_{11}$,
\boxedeq{\label{s1}
\Delta S=l^6 \displaystyle \int \left(X^{(1)}_8\swed C_3^{(0)} -\left.G_7^{(1)} \swed G_4^{(0)}+ 2G_7^{(0)}\swed G_4^{(1)}\right)\right|
~,}
where it is understood that only the bosonic $(11,0)$ components of the forms 
enter. 
This is the superinvariant corresponding to the supersymmetrization of the CS term (\ref{5.10}). The action would then read to this order,
\eq{\label{saction1}
 {S} = \int \; \Big( R^{(0)} \star 1 -\frac12 \; G^{(0)}_4 \swed \star G^{(0)}_4 - \frac16 C^{(0)}_3 \swed G^{(0)}_4 \swed G^{(0)}_4
+ l^6
\big(X^{(1)}_8\swed C_3^{(0)}  - \left.G_7^{(1)} \swed G_4^{(0)}+ 2G_7^{(0)}\swed G_4^{(1)} \big) \Big) \right|
~,}
where $R^{(0)}$, $G^{(0)}$ are identified with the fieldstrengths of the physical fields in the supergravity multiplet, while the first-order fields $R^{(1)}$, $G^{(1)}$ should be thought of as gauge-invariant functions of the physical fields. We see that the action above is in agreement with the expectation that the bosonic part of the derivative-corrected supergravity action should be of the form,
\eq{\label{saction12}
 {S} = \int \; \Big( R  \star 1 -\frac12 \; G_4 \swed \star G_4 - \frac16 C_3 \swed G_4 \swed G_4
+ l^6  \big(X_8\swed C_3+ \Delta L\star 1 \big) \Big)
~,}
with $\Delta L$ a function of $R$, $G$ and their derivatives. Since  $\Delta L$ is gauge invariant, we see in particular that the CS terms do not receive higher-order corrections beyond eight derivatives.

{Varying (\ref{saction12}) with respect to $C_3$ implies, 
\eq{\label{411}
d\star G_4+\frac12 G_4\swed G_4=X_8+\frac{\delta }{\delta C_3}(\Delta L\star 1)~.
}
It is straightforward to see that the second term on the right-hand side above is exact by virtue of the fact that $\Delta L$ 
is gauge invariant and thus only depends on $C_3$ through $G_4$. Indeed the variation of the 
$C_3$-dependent terms in the $\Delta L$ part of the action (\ref{saction12}) can be written (possibly up to integration by parts) in
 the form $\int  \Phi_7\swed\d\delta C_3 $, for some seven-form $\Phi_7$. 
Therefore by appropriately correcting the lowest-order duality relation 
by higher-derivative terms, $G_7=\star G_4+\mathcal{O}(l^6)$,  one arrives at the modified BI (\ref{gmod}).}

\subsection{How many superinvariants?}\label{sec:invariants}

We have seen  that provided the modified BI (\ref{gmod}) are satisfied, there will be at least 
one superinvariant at eight derivatives, cf. (\ref{s1}). 
A second independent superinvariant can also be similarly constructed as follows. 
Consider the twelve-form,
 \eq{
 W'_{12}= \frac16 G_4\swed G_4\swed G_4
 ~.
 \label{5.13}}
Expanding perturbatively to order $l^6$ we obtain,
\eq{
W^{\prime(1)}_{12} = \frac12 G_4^{(0)} \swed G_4^{(0)} \swed G_4^{(1)}= \d Z_{11}~;~~~ 
Z_{11}=  \frac12 G_4^{(1)} \swed G_4^{(0)}\swed C_3^{(0)}
~,}
The above can also be written in a manifestly Weil-trivial form using (\ref{gmodp}), 
\eq{
W^{\prime(1)}_{12} = \d K_{11}~;~~~ K_{11}= -G_7^{(0)}\swed G_4^{(1)}
~.}
The order-$l^6$ superinvariant corresponding  to $Z_{11}-K_{11}$ then reads,
\boxedeq{\label{s2}
\Delta S'=l^6 \displaystyle \int
G_4^{(1)} \swed\Big(\frac12 G_4^{(0)}\swed C_3^{(0)}+
G_7^{(0)}
\Big)
~,}
where it is understood that only the bosonic $(11,0)$ components of the forms 
enter. 
The above superinvariant does not contain the correct CS terms required by anomaly cancelation, cf.~(\ref{saction12}), and should therefore be excluded by 
the requirement of quantum consistency of the theory. However if one is only interested in counting superinvariants at order $l^6$ in the classical theory, the above 
superinvariant is perfectly acceptable and its existence is guaranteed provided the BI are obeyed to order $l^6$.

Dropping the requirement of quantum consistency, relying on classical supersymmetry alone, one may also consider the following two twelveforms,
\eq{\label{uvd}U_{12}=l^6 G_4\swed\mathrm{tr}R^4~;~~~
V_{12}=l^6G_4\swed(\mathrm{tr}R^2)^2
~,}
so that $U-\frac14 V$ is the Weil-trivial twelveform corresponding to the CS terms of eleven-dimensional supergravity required for anomaly cancellation, cf.~(\ref{5.10}). 
It follows that either $U$, $V$ are both Weil-trivial, or neither $U$ nor $V$ is Weil-trivial. 
If the former is true, there would exist gauge-invariant elevenforms $K_U$, $K_V$ so that at order $l^6$ we have $U^{(1)}=\d K_U$, $V^{(1)}=\d K_V$. 
One can then construct two corresponding superinvariants using the action principle,
\boxedeq{\label{suv}
\Delta S_U=l^6 \displaystyle \int \left( \mathrm{tr}R^4\swed C_3^{(0)}-K_U   \right)~;~~~
\Delta S_V=l^6 \displaystyle \int \left( (\mathrm{tr}R^2)^2\swed C_3^{(0)}-K_V\right)
~.}
By the argument at the end of the last section, $\Delta S_U$,  $\Delta S_V$ should correspond to a modified BI obtained by replacing 
the right-hand side of the second equation in (\ref{gmod}) by $\mathrm{tr}R^4$, $(\mathrm{tr}R^2)^2$ respectively. Then $K_U$, $K_V$ would still be given by (\ref{49}) but with 
$G_4^{(1)}$, $G_7^{(1)}$ solutions of the new modified BI.

Together with the superinvariant $\Delta S'$ of (\ref{s2}), we would then have a total of at least three independent superinvariants at the eight-derivative order, with only one linear 
combination thereof, $\Delta S$ of (\ref{s1}),  corresponding to the quantum-mechanically consistent eight-derivative correction. As we will see in section (\ref{sec:x8}), 
if $\Delta S_U$, $\Delta S_V$ exist they must necessarily be cubic or lower in the fields.

\subsection{$\tau$-exactness of $X_8$}\label{sec:x8}

Based on what is known about superinvariants in $D<11$ dimensions \cite{Howe:1980th}, it is plausible to 
assume that the superinvariant (\ref{s1}) corresponding to the supersymmetrization of the CS term (\ref{5.10}) should be quartic or higher in the fields.  
As pointed out  in \cite{Howe:2003cy},  a necessary condition for the superinvariant to be quartic is that the order-$l^6$ sevenform  should be quartic or higher 
in the fields. 
Since $G^{(1)}_{0,7}$  cannot be quartic or higher in the fields, as can be seen by dimensional analysis, the order-$l^6$ seven-form BI (\ref{gmodp}) must be solved for $G^{(1)}_{0,7}=0$. It then follows that the purely spinorial component of the M5-brane anomaly eightform $X_{0,8}$ is $\tau$-exact. 
Explicitly, the first nontrivial component (at dimension four) of the seven-form BI then reads,
\eq{\label{te}
\gamma^f_{(\alpha_1\alpha_2|}G^{(1)}_{f|\alpha_3\ldots\alpha_8)}=X^{(1)}_{\alpha_1\ldots\alpha_8}
~.}
As explained in detail in appendix \ref{app:ss}, taking the form of $G^{(0)}$ into account, cf.~(\ref{eq_fieldcomp}), it follows that 
the Weil-triviality condition, 
\eq{\label{wt}
W_{12}=\d K_{11}
~,
} 
is  solved up to dimension 7/2 for $K_{0,11}=K_{1,10}=K_{2,9}=0$. At dimension four, condition 
(\ref{wt}) then takes the form,
\eq{
\gamma^f_{(\alpha_1\alpha_2|}K_{fab|\alpha_3\ldots\alpha_{10})}=W^{(1)}_{ab\alpha_1\ldots\alpha_{10}}
~.}
From this it follows that (\ref{wt}) is solved,  up to $\tau$-exact terms, for $K_{3,8}$ given in terms of $G^{(1)}_{1,6}$, cf.~(\ref{te}),
\eq{\label{21}
K_{abc\alpha_1\ldots\alpha_{8}}=3\big(\gamma_{ab}\big)_{\alpha_1\alpha_2}G^{(1)}_{c\alpha_3\ldots\alpha_{8}}
~,}
where it is understood that all bosonic (spinor) indices are antisymmetrized (symmetrized). Note that the solution for $K_{3,8}$ above relies on the M2-brane identity (\ref{5.9}).

Moreover, it can be shown that all higher components of $K_{11}$ solving (\ref{wt}) are automatically guaranteed to exist. To see this, let us define the twelve-form,
\eq{I_{12}:=W_{12} 
-\d K_{11}
~,}
which is closed by construction,
\eq{\label{23}
0=
(\d I)_{p,13-p} =
\tau I_{p+1,11-p}+
\d_f   I_{p,12-p}
+\d_b   I_{p-1,13-p} 
+t  I_{p-2,14-p} 
~.}
On the other hand, as we saw above, provided (\ref{te}) holds, condition (\ref{wt}) is solved up to dimension four, i.e. $I_{p,12-p}=0$ for $p=0,1,2$. 
Setting $p=2$ in (\ref{23}) then gives $\tau I_{3,9}=0$, which  implies $I_{3,9}=0$ up to a $\tau$-exact piece that can be 
absorbed in $K_{4,7}$, since all $\tau$-cohomology groups $H^{p,12-p}_{\tau}$ are trivial 
for $p\geq3$, cf. section \ref{sec:tc}. By induction we easily see that $I_{p,12-p}=0$, for all $p\geq3$. In other words, provided 
(\ref{te}) holds, the Weil-triviality condition (\ref{wt}) is guaranteed to admit a solution.

In the present paper we provide highly nontrivial evidence corroborating (\ref{te}).  
We will give the outline of the argument here, relegating the technical details to appendix \ref{app:6}. The component 
$X_{0,8}$ of the anomaly polynomial in (\ref{te}) contains a large number of terms of the form $G^4$, which can be organized  
in terms of irreducible representations of $B_5$.  
Using certain Fierz identities, cf.~appendix \ref{app:gamma}, we have been able to show that almost all of these terms are indeed $\tau$-exact. 
There are only three irreducible representations of $B_5$ corresponding to terms which can potentially be present in $X_{0,8}$ and are not $\tau$-exact. 
These are: $(04000)$, $(03002)$ and $(02004)$, where we use the Dynkin notation for $B_5$, see e.g. appendix C of \cite{Howe:2003cy}. 

On the other hand we show that, after Fierzing, $X_{0,8}$ can be put in the form,
\eq{\spl{
X_{0,8}&=(\gamma^{a_1 a_2})(\gamma^{a_3 a_4})(\gamma^{a_5 a_6})(\gamma^{a_7 a_8}) \;  
G^4_{a_1 a_2;a_3 a_4;a_5 a_6;a_7 a_8} \\
&+(\gamma^{a_1 a_2})(\gamma^{a_3 a_4})(\gamma^{a_5 a_6})(\gamma^{a_7 \dots a_{12}}) \;   G^4_{a_1 a_2;a_3 a_4;a_5 a_6;a_7 \ldots a_{12}}   \\
&+(\gamma^{a_1 a_2})(\gamma^{a_3 a_4})(\gamma^{a_5 \dots a_9})(\gamma^{a_{10}\ldots a_{14}}) \;   G^4_{a_1 a_2;a_3 a_4;a_5 \dots a_9;a_{10} \ldots a_{14}}
~,
}}
where $G^4_{a_1 a_2;\ldots;a_7 a_8}$, $G^4_{a_1 a_2;\ldots;a_7 \ldots a_{12}}$, $G^4_{a_1 a_2;\ldots;a_{10} \ldots a_{14}}$ denote 
certain sums of $G^4$ terms with 8,4,2 indices contracted respectively, cf.~(\ref{c9}), and we have supressed spinorial indices for simplicity of notation. 
Furthermore we show that 
$(04000)$ can only be potentially present in the projection of $G^4_{a_1 a_2;\ldots;a_7 a_8}$  onto the Young diagram associated to  the partition $[4,4]$, 
while $(02004)$ can only be potentially present in the projection of $G^4_{a_1a_2;\ldots;a_{10}\ldots a_{14}}$  onto the Young diagram  $[4,4,2,2,2]$.
Therefore a necessary condition for $X_{0,8}$ to be $\tau$-exact is that the two aforementioned projections should vanish identically up to $\tau$-exact terms,
\eq{\label{rk}
\Pi_{\scalebox{0.2}{\yng(4,4)}}G^4_{a_1 a_2;\ldots;a_7 a_8}\approx0~;~~~\Pi_{\scalebox{0.2}{\yng(4,4,2,2,2)}}G^4_{a_1a_2;\ldots;a_{10}\ldots a_{14}}\approx0
~.}
In the above  $\approx$ denotes equality up to $\tau$-exact terms. 
These two constraints are highly nontrivial, involving seemingly miraculous cancellations between hundreds of terms. We have shown that, 
remarkably, (\ref{rk}) are indeed identically satisfied.

Furthermore we show  that the required cancellations for  (\ref{rk})  to hold, crucially rely on the relative coefficient between $\mathrm{tr}R^4$ and 
$(\mathrm{tr}R^2)^2$ in $X_8$. In other words we show that the purely spinorial components of $\mathrm{tr}R^4$,  
$(\mathrm{tr}R^2)^2$ are {\em not} separately $\tau$-exact. 
Consequently, if the twelve-forms $U$, $V$ are Weil-trivial, the corresponding modified order-$l^6$ BI will be solved for some $G^{(1)}_7$ which are cubic 
or lower in the fields. (Indeed if $G^{(1)}_7$ were quartic or higher, $G^{(1)}_{0,7}$ would vanish and the purely spinorial components of $\mathrm{tr}R^4$,  
$(\mathrm{tr}R^2)^2$ would be $\tau$-exact.) It then follows from (\ref{49}) that also $K_U$, $K_V$ will be cubic or lower, 
and similarly for  $\Delta S_U$, $\Delta S_V$, cf.~(\ref{suv}).

\subsection{Integrability}\label{sec:integrability}

The perturbative expansion of the curved components following from (\ref{pert}) reads,
\eq{G_{M_1\ldots M_4} = G^{(0)}_{M_1\ldots M_4} + l^6 G^{(1)}_{M_1\ldots M_4}+\cdots~,}
and similarly for $G_7$ and $R_A{}^B$. Note that in terms of 
flat components there is a mixing between zeroth order and order $l^6$ due to, 
\eq{\Phi=E^{A} \Phi_{A} = E^{(0)A} \Phi^{(0)}_A + l^6 (E^{(0)A}  \Phi^{(1)}_A + E^{(1)A} \Phi^{(0)}_A)+\cdots
~,}
where we have expanded the coframe, $E^A=E^{(0)A}+l^6E^{(1)A}+\cdots$, and we have considered an arbitrary one-form $\Phi$ for simplicity. 
However, if one restricts to the top bosonic component of a superform at $\theta=0$ as in 
(\ref{5.6}), 
then there is no mixing:
\eq{
\Phi_m^{(0)} | = e_m{}^a \Phi_a^{(0)}| +\psi_m^\alpha \Phi_\alpha^{(0)} |
~;~~~
\Phi_m^{(1)} | = e_m{}^a \Phi_a^{(1)} | +\psi_m^\alpha \Phi_\alpha^{(1)} |
~,}
where $e_m{}^a$, $\psi_m^\alpha$ were defined below (\ref{az}). 
Indeed the $\mathcal{O}(l^6)$ corrections to the coframe $E^A$ only start at higher orders in the $\theta$-expansion 
and could be systematically determined as in e.g. \cite{Tsimpis:2004gq} once the $\mathcal{O}(l^6)$ corrections to the 
torsion components have been determined.

In practice the BI are solved for the flat components of the superforms involved, $G^{(0)}_{A_1\ldots A_4} $, $G^{(1)}_{A_1\ldots A_4} $ etc, at each order in $l$. 
Consequently the corresponding BI, $\d G_4=0$ etc, are only shown to be satisfied up to terms of the next order in $l$.  In principle there may be an integrability obstruction to the 
solution of the BI at next-to-leading order in the derivative corrections, although that would most probably be prohibitively difficult to check in practice.  In the following we shall 
see that the integrability of a certain superinvariant 
is guaranteed provided the BI admit solutions to all orders in $l$. Note however that all-order integrability need not be a consequence of the BI.

The phenomenon of inducing a higher-order correction at next-to-leading order is also well understood at the level of the component action,  $S=S^{(0)} + l^6S^{(1)} +\cdots$. 
The 
condition of invariance of the action under supersymmetry transformations $\delta=\delta^{(0)} + l^6\delta^{(1)} +\cdots$  reads,
\eq{\delta^{(0)}S^{(0)}=0~;~~~
\delta^{(0)}S^{(1)}+\delta^{(1)}S^{(0)}=0~,
}
and similarly at higher orders. The term $\delta^{(1)}S^{(0)}$ in the second equation above is proportional to the lowest-order equations of motion. 
Therefore in constructing $S^{(1)}$ we only need 
to check its invariance with repsect to the lowest-order supersymmetry transformations $\delta^{(0)}$ and only up to terms which vanish by virtue of 
the lowest-order equations of motion. 
This corresponds, in the superspace approach, to the fact that  in solving the first-order BI one uses the zeroth-order equations for the various 
superfields. Once $S^{(1)}$ is thus constructed, the correction $\delta^{(1)}$ to the supersymmetry transformations can be read off. 
Since $\delta^{(1)}S^{(1)}\neq0$  in general, this induces a correction $S^{(2)}$ to the action and a corresponding correction $\delta^{(2)}$ to the supersymmetry transformations,  
and so on. 

The existence of an intergrability obstruction can also be understood in the context of the Noether procedure. Indeed at next-to-leading order 
we have,
\eq{\delta^{(2)}S^{(0)}+
\delta^{(1)}S^{(1)}+\delta^{(0)}S^{(2)}=0~.
}
Therefore there must exist an action $S^{(2)}$ such that its variation with respect to lowest-order 
supersymmetry transformations is equal to $-\delta^{(1)}S^{(1)}$, up to terms that vanish by virtue of the lowest-order equations of motion. This condition will 
not be automatically satisfied for every $S^{(1)}$.

In particular one would like to know how many of the independent superinvariants at order $l^6$ presented in section \ref{sec:invariants} survive 
to all orders in the derivative expansion. Assuming M-theory is a non-pertubatively consistent theory, we expect the superinvariant (\ref{s1}), corresponding to the 
supersymmetrization of the CS term required for anomaly cancellation, to be integrable to all orders. Moreover, assuming this superinvariant is at least quartic in the fields, 
a similar argument as the one detailed below (\ref{23}) shows that it must be unique  at order $l^6$ \cite{Howe:2003cy}.

In addition, 
if one assumes that the BI admit a solution to all orders in a perturbative expansion in $l$, then  there is one linear combination of the superinvariants presented in section  \ref{sec:invariants} that is guaranteed to exist to all orders in $l$. Indeed in that case the twelve-form,
\eq{
W_{12}= \big(
l^6X_8-\frac12 G_4\swed G_4\big)\swed G_4=\d\big(G_7\swed G_4\big)~,}
is Weil-trivial by virtue of (\ref{gmod}), which should now be considered valid to all orders in $l$. However this is {\it not} the superinvariant which corresponds to the 
supersymmetrization of the anomaly term, cf.~(\ref{s1}). Indeed  by the usual action principle procedure the twelve-form above would give rise to the superinvariant,
\eq{\label{432}\Delta S=\int\left(l^6X_8\swed C_3 - \left. \frac12 G_4\swed G_4\swed C_3- G_7\swed G_4\right)\right|
~.
}
Expanding to order $l^6$ and assuming $G_4$ receives a nonvanishing correction at this order, 
%
%
we see that (\ref{432}) does not coincide with (\ref{s1}) and the corresponding $l^6$-corrected action is different from (\ref{saction1}). 

In conclusion, under the aforementioned assumptions, we would then expect (at least) two independent superinvariants to exist to all orders in a perturbative expansion in $l$. 
Only one of these, the one corresponding to the supersymmetrization of the CS anomaly term, will be quantum-mechanically consistent.

\section{Discussion}\label{sec:disc}

We have shown that the highly nontrivial constraints (\ref{rk}) are satisfied, corroborating the expectation that the purely 
spinorial component of $X_{8}$  is $\tau$-exact. 
Furthermore we have seen that the $\tau$-exactness of $X_{0,8}$ suffices for the existence of the superinvariant at order $l^6$. 
Solving the $\tau$-exactness of $X_{0,8}$ is the first step, and arguably the most difficult, towards the explicit construction, via the action principle approach, of the supersymmetrization of the Chern-Simons term $C\wedge X_8$ of  eleven-dimensional supergravity required for the quantum consistency of the theory.

Conclusively proving the $\tau$-exactness of $X_{0,8}$ would in addition require checking that the representation $(03002)$ is absent from $X_{0,8}$. This representation is potentially present 
in two different Young diagrams. As a consequence, showing the cancellation would involve, after  projecting onto the appropriate Young diagram, Fierzing hundreds of four-$\gamma$ terms. This is equivalent to eight-spinor Fierzing, as opposed to the four-spinor Fierzing which is  sufficient in order to show the absence of the $(04000)$ and $(02004)$ representations. At present, this 
seems prohibitively difficult even with the help of a computer.

As a corollary of this work, we have shown that if  the anomalous Chern-Simons terms $C\wedge  (\mathrm{Tr}R^2)^2$ and $C\wedge \mathrm{Tr}R^4$ can be supersymmetrized independently, the corresponding superinvariants must necessarily contain terms cubic or lower in the fields. The existence of eleven-dimensional cubic superinvariants at the eight-derivative order has not 
been examined in the past. Their existence would presumably imply, by dimensional reduction, the presence of cubic terms in the ten-dimensional superinvariants $I_{0a}$, $I_{0b}$ mentioned in the introduction. This would not  be inconsistent with the results of \cite{deRoo:1993} who have excluded from the outset such terms in 
their analysis. This is an interesting 
open question to which we hope to return in the future.

\section*{Acknowledgements}
We would like to thank Paul Howe for valuable discussions.~D.T. would  like to thank the Galileo Galilei Institute for Theoretical Physics for  hospitality and the INFN for partial support during the completion of this work.

\appendix

\section{Weil triviality at $l^0$}\label{app:0}

In this section we give the details of the solution of the superspace equation $W_{12}=\d K_{11}$ at lowest order in the Planck length. As 
a byproduct we will see that the solution for $K_{11}$ given in  (\ref{ty}) is unique up to exact terms. 
We will look for the solution to,
\begin{align}
\d K_{11} =  -\frac{1}{6} \;G_4 \swed G_4 \swed G_4~,
\label{eq_KL0}
\end{align}
for $K_{11}$ gauge-invariant, i.e. function of the fieldstrengths of the physical fields. 
The explicit construction of $K_{11}$ in flat components proceeds by solving the BI at each engineering dimension in a stepwise fashion, from dimension $-3$ to $2$ (i.e. from  $K_{\alpha_1 \dots \alpha_{11}}$ to $K_{a_1 \dots a_{11}}$). In components the BI (\ref{eq_KL0}) reads,
%
%
\begin{align}
D_{[A_{1}}  K_{A_{2} \dots A_{12})} + \frac{11}{2} \; T_{[A_{1} A_{2}|}^{\qquad F} \; K_{F|A_{3} \dots A_{12})} = -\frac{11!}{6(4!)^3} \; G_{[A_{1} \dots A_{4}} \; G_{A_{5} \dots A_{8}} \; G_{A_{9} \dots A_{12})},
\label{eq_dKeqGcompL0}
\end{align}
where the torsion term arises from the the action of the exterior derivative on the supervielbein. The $[A B C)$ notation stands for symmetrization or antisymmetrization, depending on the bosonic or fermionic nature of the indices. {\em In the following,  antisymmetrisation of the indices $a_i$ and symmetrisation of the indices $\alpha_i$ is always implied.}

The engineering (mass) dimensions of the physical fields which will be involved in the construction of $K_{11}$ are,
\vspace{-1cm}
\begin{minipage}[t]{0.15\linewidth}
\begin{align*}
\left[ D_{a_1} \right] &= 1 \\
\left[ D_{\alpha_1} \right] &= 1/2
\end{align*}
\end{minipage}
\hfill
\begin{minipage}[t]{0.15\linewidth}
\begin{align*}
\left[ {T_{a_1 a_2}}^{\alpha} \right] &= 3/2 \\
\left[ {T_{a \alpha}}^{\beta} \right] &= 1
\end{align*}
\end{minipage}
\hfill
\begin{minipage}[t]{0.15\linewidth}
\begin{align*}
\left[ G_{a b \alpha \beta} \right] &= \left[ {T_{\alpha \beta}}^{a} \right] = 0 \notag \\
\left[ G_{abcd} \right] &= 1 \notag
\end{align*}
\end{minipage}
\hfill
\begin{minipage}[t]{0.1\linewidth}
\begin{align*}
\hphantom{a}
\end{align*}
\end{minipage}
\\
\\

\subsection*{From dimension -3 to -1/2}

From dimension $-3$ ($12$ odd indices) to $-1/2$ ($7$ odd and $5$ even indices), the right hand side of (\ref{eq_KL0}) always vanishes. Given the dimensions of the fieldstrengths of the physical fields, the first non-vanishing component of $K_{11}$ is $K_{\alpha_1 \dots \alpha_4 a_1 \dots a_7}$, appearing for the first time in the $0$-dimensional equation ($6$ fermionic indices and $6$ bosonic indices). For example, the equation (\ref{eq_dKeqGcompL0}) at dimension $-1/2$ reads,
\begin{align*}
& \frac{7}{12} \; D_{\alpha_{1}} K_{\alpha_{2} \dots \alpha_{7} a_{1} \dots a_{5}} \;-\; \frac{5}{12} D_{a_{1}} K_{a_{2} \dots a_{5} \alpha_{1} \dots \alpha_{7}} \;+ \\
& \frac{11}{2} \bigg( \frac{5}{33} \; {T_{\alpha_{1} \alpha_{2}}}^{f} \; K_{f \alpha_{3} \dots \alpha_{7} a_{1} \dots a_{5}} \;-\; \frac{7}{22} \; {T_{a_{1} a_{2}}}^{\gamma} \; K_{\gamma a_{3} \dots a_{5} \alpha_{1} \dots \alpha_{7}} \;-\; \frac{35}{66} \; {T_{a_{1} \alpha_{1}}}^{\gamma} \; K_{\gamma \alpha_{2} \dots \alpha_{7} a_{2} \dots a_{5}} \bigg) = 0 ~,
\end{align*}
and involves\footnote{In the following we will use superscripts to indicate the dimension. This should not be confused with the notation in the main text, e.g.~(\ref{pert}) 
where the superscript denotes the order in the derivative expansion.} $K_{\alpha_1 \dots \alpha_5 a_1 \dots a_6}^{(-1/2)}$, $K_{\alpha_1 \dots \alpha_6 a_1 \dots a_5}^{(-1)}$, $K_{\alpha_1 \dots \alpha_7 a_1 \dots a_4}^{(-3/2)}$ and $K_{\alpha_1 \dots \alpha_8 a_1 \dots a_3}^{(-2)}$, which cannot be expressed in terms of the physical fields: the equation is thus trivially satisfied.

\subsection*{Dimension 0 - ($A_{1} \dots A_{6} \rightarrow \alpha_1 \dots \alpha_{6}$, $A_{7} \dots A_{12} \rightarrow a_1 \dots a_{6}$)}

At dimension $0$, eq.~(\ref{eq_dKeqGcompL0}) reads:
\begin{align*}
& \frac{1}{2} \; D_{\alpha_{1}} \overbrace{K_{\alpha_{2} \dots \alpha_{6} a_{1} \dots a_{6}}}^{0} \;+\; \frac{1}{2} D_{a_{1}} \overbrace{K_{a_{2} \dots a_{6} \alpha_{1} \dots \alpha_{6}}}^{0} \;+ \\
& \frac{11}{2} \bigg( \frac{5}{22} \; {T_{\alpha_{1} \alpha_{2}}}^{f} \; K_{f \alpha_{3} \dots \alpha_{6} a_{1} \dots a_{6}} \;+\; \frac{5}{22} \; {T_{a_{1} a_{2}}}^{\gamma} \; \underbrace{K_{\gamma a_{3} \dots a_{6} \alpha_{1} \dots \alpha_{6}}}_{0} \;+\; \frac{12}{22} \; {T_{a_{1} \alpha_{1}}}^{\gamma} \; \underbrace{K_{\gamma \alpha_{2} \dots \alpha_{6} a_{2} \dots a_{6}}}_{0} \bigg) \\ 
&= -\frac{11!}{6(4!)^3} \; \frac{18}{77} \; G_{a_1 a_2 \alpha_1 \alpha_2} G_{a_3 a_4 \alpha_3 \alpha_4} G_{a_5 a_6 \alpha_5 \alpha_6}~.
\end{align*}
Most terms vanish and the equation simplifies as follows,
\begin{align*}
(\gamma^{f})_{\alpha_1 \alpha_2} K_{f a_1 \dots a_6 \alpha_3 \dots \alpha 6}
= 90 \; (\gamma_{a_1 a_2})_{\alpha_1 \alpha_2} (\gamma_{a_3 a_4})_{\alpha_3 \alpha_4} (\gamma_{a_5 a_6})_{\alpha_5 \alpha_6}.
\end{align*}
Using the M2-brane identity as well as the so-called M5-brane identity, 
\eq{\label{m5bi}
(\gamma^{e})_{\alpha_1 \alpha_2} (\gamma_{e a_1 \dots a_4})_{\alpha_3\alpha_4} = 3 \; (\gamma_{a_1 a_2})_{\alpha_1 \alpha_2} (\gamma_{a_3 a_4})_{\alpha_3 \alpha_4}~, 
}
it is easy to check that the solution is given by,
\begin{align}
K_{a_1 \dots a_7 \alpha_1 \dots \alpha 4}
= 42 \; (\gamma_{a_1 \dots a_5})_{\alpha_1 \alpha_2} (\gamma_{a_6 a_7})_{\alpha_3 \alpha_4}~.
\label{eq_exprKd0L0}
\end{align}

\subsection*{Dimension 1/2 - ($A_{1} \dots A_{5} \rightarrow \alpha_1 \dots \alpha_{5}$, $A_{6} \dots A_{12} \rightarrow a_1 \dots a_{7}$)}

At dimension $1/2$, eq.~(\ref{eq_dKeqGcompL0}) reads,
\begin{align*}
& \frac{5}{12} \; \overbrace{D_{\alpha_{1}} K_{\alpha_{2} \dots \alpha_{5} a_{1} \dots a_{7}}}^{D_{\alpha} \left( \gamma^{(5)} \gamma^{(2)} \right) = 0} \;-\; \frac{7}{12} \; D_{a_{1}} \overbrace{K_{a_{2} \dots a_{7} \alpha_{1} \dots \alpha_{5}}}^{0} \;+ \\
& \frac{11}{2} \bigg( \frac{5}{33} \; {T_{\alpha_{1} \alpha_{2}}}^{f} \; K_{f \alpha_{3} \dots \alpha_{5} a_{1} \dots a_{7}} \;-\; \frac{35}{66} \; {T_{a_{1} a_{2}}}^{\gamma} \; \underbrace{K_{\gamma a_{3} \dots a_{7} \alpha_{1} \dots \alpha_{5}}}_{0} \;-\; \frac{7}{22} \; {T_{a_{1} \alpha_{1}}}^{\gamma} \; \underbrace{K_{\gamma \alpha_{2} \dots \alpha_{6} a_{2} \dots a_{6}}}_{0} \bigg) = 0~,
\end{align*}
which simplifies to,
\begin{align*}
(\gamma^{f})_{\alpha_1 \alpha_2} K_{f a_1 \dots a_7 \alpha_3 \dots \alpha_5} = 0~.
\end{align*}
Since $\left[ K_{f a_1 \dots a_7 \alpha_3 \dots \alpha_5} \right] = 1/2$ and there is no gauge-invariant field with that dimension, we conclude,
\begin{align}
K_{a_1 \dots a_8 \alpha_1 \dots \alpha_3} = 0~.
\label{eq_exprKd12L0}
\end{align}

\subsection*{Dimension 1 - ($A_{1} \dots A_{4} \rightarrow \alpha_1 \dots \alpha_{4}$, $A_{5} \dots A_{12} \rightarrow a_1 \dots a_{8}$)}

At dimension $1$, eq.~(\ref{eq_dKeqGcompL0}) reads,
\begin{align*}
& \frac{4}{12} \; D_{\alpha_{1}} \overbrace{K_{\alpha_{2} \dots \alpha_{4} a_{1} \dots a_{8}}}^{0} \;+\; \frac{8}{12} \; \overbrace{D_{a_{1}} K_{a_{2} \dots a_{8} \alpha_{1} \dots \alpha_{4}}}^{\d_{a_1} \left( \gamma^{(5)} \gamma^{(2)} \right) = 0} \;+ \\
& \frac{11}{2} \bigg( \frac{1}{11} \; {T_{\alpha_{1} \alpha_{2}}}^{f} \; K_{f \alpha_{3} \alpha_{4} a_{1} \dots a_{8}} \;+\; \frac{14}{33} \; {T_{a_{1} a_{2}}}^{\gamma} \; \underbrace{K_{\gamma a_{3} \dots a_{8} \alpha_{1} \dots \alpha_{4}}}_{0} \;+\; \frac{16}{33} \; {T_{a_{1} \alpha_{1}}}^{\gamma} \; K_{\gamma \alpha_{2} \dots \alpha_{4} a_{2} \dots a_{8}} \bigg) \\ 
&= -\frac{11!}{6(4!)^3} \; \frac{12}{55} \; G_{a_1 a_2 a_3 a_4} G_{a_5 a_6 \alpha_1 \alpha_2} G_{a_7 a_8 \alpha_3 \alpha_4}~,
\end{align*}
which becomes, using (\ref{eq_fieldcomp}),
\begin{align}
(\gamma^{f})_{\alpha_1 \alpha_2} K_{f a_1 \dots a_8 \alpha_3 \alpha_4} =& 
- \frac{56}{3} \; i \; G_{a_1 a_2 a_3 f} \; (\gamma^{f})_{\alpha_1 \alpha_2} \; (\gamma_{a_4 \dots a_8})_{\alpha_3 \alpha_4}  \notag \\
&+ \frac{7}{18} \; i \; G_{f g h i} \; ({\gamma_{a_1 \dots a_6}}^{fghi})_{\alpha_1 \alpha_2} \; ({\gamma_{a_7 a_8}})_{\alpha_3 \alpha_4} \notag \\
&+ 70 \; i \; (\gamma_{a_1 a_2})_{\alpha_1 \alpha_2} (\gamma_{a_3 a_4})_{\alpha_3 \alpha_4} G_{a_5 \dots a_8}~.
\label{eq_Kd1L0}
\end{align}
The last term above can be expanded as,
\begin{align*}
70 \;i\; (\gamma_{a_1 a_2})_{\alpha_1 \alpha_2} \; & (\gamma_{a_3 a_4})_{\alpha_3 \alpha_4} \; G_{a_5 \dots a_8}
= \frac{70}{3} \;i\; (\gamma^{f})_{\alpha_1 \alpha_2} \; (\gamma_{f a_1 \dots a_4})_{\alpha_3 \alpha_4} \; G_{a_5 \dots a_8} \\
&= 42 \;i\; (\gamma^{f})_{\alpha_1 \alpha_2} \; (\gamma_{[f a_1 \dots a_4|})_{\alpha_3 \alpha_4} \; G_{|a_5 \dots a_8]} - \frac{56}{3} \;i\; (\gamma^{f})_{\alpha_1 \alpha_2} \; (\gamma_{[a_1 \dots a_5|})_{\alpha_3 \alpha_4} \; G_{|a_6 a_7 a_8] f}~.
\end{align*}
Similarly, the second term on the right-hand side of (\ref{eq_Kd1L0}) can be written in a manifestly $\tau$-exact form,
\begin{align*}
\frac{7}{18} \; ({\gamma_{a_1 \dots a_6}}^{fghi})_{\alpha_1 \alpha_2} \; & (\gamma_{a_7 a_8})_{\alpha_3 \alpha_4} 
= -\frac{7}{18} \; {\epsilon_{j a_1 \dots a_6}}^{fghi} \; (\gamma^{j})_{\alpha_1 \alpha_2} \; (\gamma_{a_7 a_8})_{\alpha_3 \alpha_4}  \\
&= -\frac{1}{2} \; {\epsilon_{[j a_1 \dots a_6|}}^{fghi} \; (\gamma^{j})_{\alpha_1 \alpha_2} \; (\gamma_{|a_7 a_8]})_{\alpha_3 \alpha_4} + \frac{1}{9} \; {\epsilon_{a_1 \dots a_7}}^{fghi} \; \underbrace{(\gamma^{j})_{\alpha_1 \alpha_2} \; (\gamma_{a_8 j})_{\alpha_3 \alpha_4}}_{0}~.
\end{align*}
Then eq.~(\ref{eq_Kd1L0}) takes the following form,
\begin{align*}
(\gamma^{j})_{\alpha_1 \alpha_2} K_{j a_1 \dots a_8 \alpha_3 \alpha_4} = (\gamma^{j})_{\alpha_1 \alpha_2} \Big(42 \; i \; (\gamma_{[j a_1 \dots a_4})_{\alpha_3 \alpha_4} \; G_{a_5 \dots a_8]} - \frac{1}{2} \; i \; {\epsilon_{[j a_1 \dots a_6|}}^{i_1 \dots i_4} \; (\gamma_{|a_7 a_8]})_{\alpha_3 \alpha_4} \; G_{i_1 \dots i_4} \Big)~.
\end{align*}
Since the cohomology group $H^{9,2}_{\tau}$ is trivial, the  solution to the above equation reads,
\begin{align*}
K_{a_1 \dots a_9 \alpha_1 \alpha_2} = 42 \; i \; (\gamma_{a_1 \dots a_5})_{\alpha_1 \alpha_2} \; G_{a_6 \dots a_9} - \frac{1}{2} \;i\; {\epsilon_{a_1 \dots a_7}}^{i_1 \dots i_4} \; (\gamma_{a_8 a_9})_{\alpha_1 \alpha_2} \; G_{i_1 \dots i_4}~,
\end{align*}
up to $\tau$-exact terms.

\subsection*{Dimension 3/2 - ($A_{1} \dots A_{3} \rightarrow \alpha_1 \dots \alpha_{3}$, $A_{4} \dots A_{12} \rightarrow a_1 \dots a_{9}$)}

At dimension $3/2$, eq.~(\ref{eq_dKeqGcompL0}) reads,
\begin{align*}
& \frac{3}{12} \; D_{\alpha_{1}} K_{\alpha_{2} \alpha_{3} a_{1} \dots a_{9}} \;-\; \frac{9}{12} D_{a_{1}} \overbrace{K_{a_{2} \dots a_{9} \alpha_{1} \dots \alpha_{3}}}^{0} \;- \\
& \frac{11}{2} \bigg( \frac{1}{22} \; {T_{\alpha_{1} \alpha_{2}}}^{f} \; K_{f a_{1} \dots a_{9} \alpha_{3}} \;+\; \frac{6}{11} \; {T_{a_{1} a_{2}}}^{\gamma} \; K_{\gamma a_{3} \dots a_{9} \alpha_{1} \dots \alpha_{3}} \;+\; \frac{9}{22} \; {T_{a_{1} \alpha_{1}}}^{\gamma} \; \underbrace{K_{\gamma \alpha_{2} \alpha_{3} a_{2} \dots a_{9}}}_{0} \bigg) = 0~,
\end{align*}
which becomes, using (\ref{eq_fieldcomp}),
\begin{align}
(\gamma^{f})_{\alpha_1 \alpha_2} K_{f a_1 \dots a_9 \alpha_3} =
&+ 252 \; (\gamma_{a_1 \dots a_5})_{\alpha_2 \alpha_3} \; (\gamma_{a_6 a_7})_{\alpha_1 \gamma} \; {T_{a_8 a_9}}^{\gamma} \notag \\
&- 3 \; {\epsilon_{a_1 \dots a_7}}^{i_1 \dots i_4} (\gamma_{a_8 a_9})_{\alpha_1 \alpha_2} \; (\gamma_{i_1 i_2})_{\alpha_3 \gamma} \; {T_{i_3 i_4}}^{\gamma} \notag \\
&+ 504 \; (\gamma_{a_1 a_2})_{(\alpha_1 \alpha_2|} \; (\gamma_{a_3 \dots a_7})_{|\alpha_3 \gamma)} \; {T_{a_8 a_9}}^{\gamma}~.
\label{eq_Kd32L0}
\end{align}
The decomposition of $K_{f a_1 \dots a_9 \alpha_3}$ in irreducible components is given by $$(10000) \otimes (00001) = (10001) \oplus (00001)~,$$ whereas ${T_{ab}}^{\alpha}$ is in the representation $(01001)$. It follows that,
\begin{align}
K_{a_1 \dots a_{10} \alpha_1} = 0~,
\end{align}
and moreover the right-hand side of (\ref{eq_Kd32L0}) must vanish identically. This  can be verified by e.g. taking the Hodge dual of $(\gamma_{i_1 i_2})_{\alpha_3 \gamma}$ in the second term of (\ref{eq_Kd32L0}), and using the $\gamma$-tracelessness of $T_{ab}{}^{\gamma}$, cf.~(\ref{eq_eom}).

\subsection*{Dimension 2 - ($A_{1} A_{2} \rightarrow \alpha_1 \alpha_{2}$, $A_{3} \dots A_{12} \rightarrow a_1 \dots a_{10}$)}

At dimension $2$, eq.~(\ref{eq_dKeqGcompL0}) reads,
\begin{align*}
& \frac{2}{12} \; D_{\alpha_{1}} K_{\alpha_{2} a_{1} \dots a_{10}} \;+\; \frac{10}{12} D_{a_{1}} K_{a_{2} \dots a_{10} \alpha_{1} \alpha_{2}} \;+ \\& \frac{11}{2} \bigg( \frac{1}{66} \; {T_{\alpha_{1} \alpha_{2}}}^{f} \; K_{f a_{1} \dots a_{10}} \;+\; \frac{10}{33} \; {T_{a_{1} a_{2}}}^{\gamma} \; K_{\gamma a_{3} \dots a_{10} \alpha_{1} \alpha_{2}} \;+\; \frac{15}{22} \; {T_{a_{1} \alpha_{1}}}^{\gamma} \; K_{\gamma \alpha_{2} a_{2} \dots a_{10}} \bigg) \\ 
&= -\frac{11!}{6(4!)^3} \; \frac{18}{66} \; G_{a_1 a_2 a_3 a_4} G_{a_5 a_6 a_7 a_8} G_{a_9 a_{10} \alpha_1 \alpha_2}~,
\end{align*}
which becomes, using (\ref{eq_fieldcomp}),
\begin{align}
&(\gamma^{f})_{\alpha_1 \alpha_2} K_{f a_1 \dots a_{10}} = \notag \\
&-10 \;i \left(42 \; i \; (\gamma_{a_2 \dots a_6})_{\alpha_1 \alpha_2} \; \d_{a_1} G_{a_7 \dots a_{10}} - \frac{1}{2} \;i\; {\epsilon_{a_2 \dots a_8}}^{i_1 \dots i_4} \; (\gamma_{a_9 a_{10}})_{\alpha_1 \alpha_2} \; \d_{a_1} G_{i_1 \dots i_4} \right) \notag \\
&-20 \;i\; {T_{a_1 \alpha_1}}^{\epsilon} \Big( 42 \; i \; (\gamma_{a_2 \dots a_6})_{\epsilon \alpha_2} \; G_{a_7 \dots a_{10}} - \frac{1}{2} \;i\; {\epsilon_{a_2 \dots a_8}}^{i_1 \dots i_4} \; (\gamma_{a_9 a_{10}})_{\epsilon \alpha_2} \; G_{i_1 \dots i_4} \Big) \notag \\
&- 1575 \; G_{a_1 \dots a_4} G_{a_5 \dots a_8} (\gamma_{a_9 a_{10}})_{\alpha_1 \alpha_2}~.
\label{eq_Kd2L0}
\end{align}
Muliplying by $\gamma^{(1)}$ and taking the trace leads to,
\begin{align}
K_{a_1 \dots a_{11}} = \frac{1}{72} \; \epsilon_{a_1 \dots a_{11}} G_{\d_1 \dots \d_4} G^{\d_1 \dots \d_4}~.
\label{eq_exprKd2L0}
\end{align}
On the other hand contracting (\ref{eq_Kd2L0}) with $\gamma^{(2)}$ or $\gamma^{(5)}$  imposes that the contraction of the right-hand side must be identically zero. 
This can indeed be straightforwardly  verified using (\ref{eq_eom}). 

\subsection*{Dimension 5/2 - ($A_{1} \rightarrow \alpha_1$, $A_{2} \dots A_{12} \rightarrow a_1 \dots a_{11}$)}

The equation at dimension $5/2$ does not contain any additional information, but serves as a consistency check for the expressions we found for $K_{a_1 \dots a_{11}}$. It reads,
\begin{align*}
& \frac{1}{12} \; D_{\alpha_{1}} K_{a_{1} \dots a_{11}} \;-\; \frac{11}{12} \; D_{a_{1}} K_{a_{2} \dots a_{11} \alpha_{1}} \;- \frac{11}{2} \bigg( \frac{2}{12} \; {T_{a_{1} a_{2}}}^{\gamma} \; K_{\gamma a_{3} \dots a_{11} \alpha_{1}} \;-\; \frac{10}{12} \; {T_{a_{1} \alpha_{1}}}^{\gamma} \; K_{\gamma a_{2} \dots a_{11}} \bigg) = 0~,
\end{align*}
which becomes, using (\ref{eq_fieldcomp}) and (\ref{eq_exprKd2L0}),
\begin{align*}
\frac{1}{72} \; \epsilon_{a_1 \dots a_{11}} \; D_{\alpha_1} G_{abcd} G^{abcd}
- 330 \;i\; G_{a_1 a_2 g h} \; ({\gamma^{gh}}_{a_3 \dots a_{9}} \; T_{a_{10} a_{11}})_{\alpha_1}
+ 2310 \;i\; G_{a_1 \dots a_4} \; (\gamma_{a_5 \dots a_{9}} \; T_{a_{10} a_{11}})_{\alpha_1}~.
\end{align*}
Using (\ref{eq_spinder}), (\ref{eq_hodge}) we then obtain the constraint,
\begin{align}
0 = \epsilon_{a_1 \dots a_{11}} \; {T_{d_1 d_2}}^{\delta} \; (\gamma_{d_3 d_4})_{\delta \alpha_1} \; G^{d_1 \dots d_4} 
&+ \frac{77}{4} \; {\epsilon_{a_1 \dots a_5}}^{b_1 \dots b_6} \; (\gamma_{b_1 \dots b_6} \; T_{a_6 a_7})_{\alpha_1} \; G_{a_8 \dots a_{11}} \notag \\
&- 990 \; {\epsilon_{a_1 \dots a_7}}^{b_1 b_2 g h} \; (\gamma_{b_1 b_2} \; T_{a_8 a_9})_{\alpha_1} \; G_{a_{10} a_{11} g h}~,
\label{eq_Kd52L0} 
\end{align}
which can be seen to be automatically satisfied by contracting (\ref{eq_Kd52L0}) with $\epsilon_{a_1 \dots a_{11}}$. The next equation (of dimension 3) is trivially satisfied, since the  purely bosonic component of a twelveform  vanishes automatically in eleven dimensions.

\subsection*{Action at $\mathcal{O}(l^0)$}

We have thus constructed the explicit expression of all components of $K_{11}$ and have seen that it is unique up to exact terms. 
Its purely bosonic component in particular takes the following form,
\begin{align}\label{br}
K^{(2)} = \frac{1}{72 \cdot 11!} \; \epsilon_{a_1 \dots a_{11}} G_{d_1 \dots d_4} \; G^{d_1 \dots d_4} \; dx^{a_1} \swed \dots \swed dx^{a_{11}} &= \frac{1}{3} \; G \swed \star G \\
&= - \; R \star 1 +\frac12 \; G \swed \star G~, 
\end{align}
where in this subsection we have reverted to bosonic conventions for bosonic forms. 
Using the action principle then leads to the CJS action of section \ref{sec:cjs}.

The last two equalities in (\ref{br}) above can be seen as follows. 
The volume element is defined as, $$\d V = \star 1 = \frac{1}{11!} \; \epsilon_{a_1 \dots a_{11}} \; dx^{a_1} \swed \dots \swed dx^{a_{11}}~,$$ 
from which it follows that,
\begin{align*}
G \swed \star G
\;=\; \frac{1}{7! \; (4!)^2} \; G_{a_1 \dots a_4} \; {\epsilon_{a_5 \dots a_{11}}}^{b_1 \dots b_4} \; G_{b_1 \dots b_4} \; \overbrace{dx^{a_1} \swed \dots \swed dx^{a_{11}}}^{- \epsilon^{a_1 \dots a_{11}} \d V} 
\;=\; \frac{1}{4!} \; G_{d_1 \dots d_4} \; G^{d_1 \dots d_4} \; \d V~.
\end{align*}
Moreover, taking the trace of  the third relation of (\ref{eq_eom}) gives,
\begin{align*}
R \star 1 = \frac{1}{144} \; G_{d_1 \dots d_4} G^{d_1 \dots d_4} \; \d V = \frac{1}{6} \; G \swed \star G~.
\end{align*}

\section{Weil triviality at $l^3$}\label{app:3}

In this section we are looking for the solution to the equation,
\begin{align}
\d K_{11} = G_4 \swed G_4 \swed R^{ab} \swed R_{ba}~.
\end{align}
We will construct all components of $K_{11}$ explicitly and confirm that the solution of section \ref{sec:five} is unique up to exact terms. 
In components the equation above  takes the following form,
\begin{align}
D_{[A_{1}} \; K_{A_{2} \dots A_{12})} + \frac{11}{2} \; T_{[A_{1} A_{2}|}^{\qquad F} \; K_{F|A_{3} \dots A_{12})} = \frac{11!}{4(4!)^2} \; R_{[A_{1} A_{2} | c_1 c_2} \; {R_{| A_{3} A_{4}|}}^{c_2 c_1} \; G_{|A_{5} \dots A_{8}} \; G_{A_{9} \dots A_{12})}~.
\label{eq_dKeqGcompL3}
\end{align}
The dimensions of the physical fields are the same as before, with the addition of $\left[ R_{abcd} \right] = 2$. The dimensions of the various components of $K$ range from $-1/2$ ($K_{\alpha_1 \dots \alpha_{11}}$) to $5$ ($K_{a_1 \dots a_{11}}$).

\subsection*{Dimension 0 to 3/2}

Since the dimension of $K_{\alpha_1 \dots \alpha_{11}}$ is $-1/2$, it must be set to zero as it cannot be expressed in terms of the physical fields. 
The equation of dimension $0$ then takes the form,
\begin{align*}
D_{\alpha_{1}} \underbrace{K_{\alpha_{2} \dots \alpha_{12}}}_{0} \;+\; \frac{11}{2} \; {T_{\alpha_{1} \alpha_{2}}}^{f} \; K_{f \alpha_{3} \dots \alpha_{12}}
= \frac{11!}{4(4!)^2} \; {R_{\alpha_1 \alpha_2}}_{c_1 c_2} \; {R_{\alpha_3 \alpha_4}}^{c_2 c_1} \; \underbrace{G_{\alpha_5 \dots \alpha_8} \; G_{\alpha_9 \dots \alpha_{12}}}_{0}~,
\end{align*}
which simplifies to,
\begin{align}
(\gamma^{f})_{\alpha_{1} \alpha_{2}} \; K_{f \alpha_{3} \dots \alpha_{12}} = 0~.
\label{eq_Kd0L3}
\end{align}
Since $\left[K_{f \alpha_{3} \dots \alpha_{12}}\right]=0$ and $H_{\tau}^{1,10}$ is nontrivial, a $\tau$-nonexact solution involving only $\gamma$-matrices could exist. In that case 
$K_{f \alpha_{3} \dots \alpha_{12}}$ would necessary transform as a scalar, since the only available gauge-invariant superfield of zero dimension is a constant. On the other hand,  
$$(10000)\otimes(00001)^{\otimes_S 10}=1\times (00000)+\cdots~,$$ 
i.e. the decomposition of $K_{f \alpha_{3} \dots \alpha_{12}}$ contains a unique scalar combination. It follows that,   
\eq{K_{f \alpha_{3} \dots \alpha_{12}}\propto
 (\gamma_{f})_{\alpha_3 \alpha_4} (\gamma^{a})_{\alpha_5 \alpha_6} (\gamma_{a})_{\alpha_7 \alpha_8} (\gamma^{b})_{\alpha_9 \alpha_{10}}
 (\gamma_{b})_{\alpha_{11} \alpha_{12}}~.
}
However it can be verified that this expression does not satisfy eq.~(\ref{eq_Kd0L3}), unless $K_{a_1 \alpha_3 \dots \alpha_{12}} = 0$.

The right-hand side of eq.~(\ref{eq_dKeqGcompL3}) vanishes from dimension $0$ to dimension $3/2$, and the equations to solve are all similar to (\ref{eq_Kd0L3}): The component $K^{(1/2)}_{a_1 a_2 \alpha_{1} \dots \alpha_{9}}$ will be set to zero because there is no gauge-invariant field of dimension $1/2$. The components $K^{(1)}_{a_1 a_2 a_3 \alpha_{1} \dots \alpha_{8}}$, $K^{(3/2)}_{a_1 a_2 a_3 a_4 \alpha_{1} \dots \alpha_{7}}$ will be set to zero, up to exact terms, as a consequence of the triviality of $H^{3,8}_{\tau}$, $H^{4,7}_{\tau}$.

\subsection*{Dimension 2 - ($A_{1} \dots A_{8} \rightarrow \alpha_1 \dots \alpha_{8}$, $A_{9} \dots A_{12} \rightarrow a_1 \dots a_{4}$)}

This is the first equation with a non-zero right-hand side,
\begin{align*}
& \frac{8}{12} \; D_{\alpha_{1}} \overbrace{K_{\alpha_{2} \dots \alpha_{8} a_{1} \dots a_{4}}}^{0} \;+\; \frac{4}{12} \; D_{a_{1}} \overbrace{K_{a_{2} \dots a_{4} \alpha_{1} \dots \alpha_{8}}}^{0} \\
+ &\frac{11}{2} \bigg( \frac{14}{33} \; {T_{\alpha_{1} \alpha_{2}}}^{f} \; K_{f a_{1} \dots a_{4} \alpha_{3} \dots \alpha_8} \;+\; \frac{1}{11} \; {T_{a_{1} a_{2}}}^{\gamma} \; \underbrace{K_{\gamma \alpha_{1} \dots \alpha_{8} a_{3} a_{4}}}_{0} \;+\; \frac{16}{33} \; {T_{a_{1} \alpha_{1}}}^{\gamma} \; \underbrace{K_{\gamma \alpha_{3} \dots \alpha_{8} a_{2} \dots a_{4}}}_{0} \bigg) \\
= &3 \; \frac{4}{55} \; \frac{11!}{4(4!)^2} \; R_{\alpha_1 \alpha_2 c_1 c_2} \; {R_{\alpha_3 \alpha_4}}^{c_2 c_1} \; G_{a_1 a_2 \alpha_5 \alpha_6} \; G_{a_3 a_4 \alpha_7 \alpha_8}~,
\end{align*}
which becomes, using (\ref{eq_fieldcomp}),
\begin{align*}
(\gamma^{f})_{\alpha_1 \alpha_2} \; K_{f a_{1} \dots a_{4} \alpha_{3} \dots \alpha_8} = - 180 \;i\; (\gamma^{f})_{\alpha_1 \alpha_2} (\gamma_{f a_1 \dots a_4})_{\alpha_3 \alpha_4} \; {R_{\alpha_5 \alpha_6}}^{c_1 c_2} \; R_{\alpha_7 \alpha_8 c_2 c_1}~.
\end{align*}
Since $H_{\tau}^{5,6}$ is trivial, the solution reads,
\begin{align*}
K^{(2)}_{a_{1} \dots a_{5} \alpha_{1} \dots \alpha_6} = -180 \;i\; (\gamma_{a_1 \dots a_5})_{\alpha_1 \alpha_2} \; {R_{\alpha_3 \alpha_4}}^{c_1 c_2} R_{\alpha_5 \alpha_6 c_2 c_1}~,
\end{align*}
up to $\tau$-exact terms.

\subsection*{Dimension 5/2 - ($A_{1} \dots A_{7} \rightarrow \alpha_1 \dots \alpha_{7}$, $A_{8} \dots A_{12} \rightarrow a_1 \dots a_{5}$)}

At dimension $5/2$, eq.~(\ref{eq_dKeqGcompL3}) reads,
\begin{align*}
& \frac{7}{12} \; D_{\alpha_{1}} K_{\alpha_{2} \dots \alpha_{7} a_{1} \dots a_{5}} \;-\; \frac{5}{12} \; D_{a_{1}} \overbrace{K_{a_{2} \dots a_{5} \alpha_{1} \dots \alpha_{7}}}^{0} \\
- &\frac{11}{2} \Big( \frac{7}{22} \; {T_{\alpha_{1} \alpha_{2}}}^{f} \; K_{f a_{1} \dots a_{5} \alpha_{3} \dots \alpha_7} \;+\; \frac{5}{33} \; {T_{a_{1} a_{2}}}^{\gamma} \; \underbrace{K_{\gamma \alpha_{1} \dots \alpha_{7} a_{3} \dots a_{5}}}_{0} \;+\; \frac{35}{66} \; {T_{a_{1} \alpha_{1}}}^{\gamma} \; \underbrace{K_{\gamma \alpha_{3} \dots \alpha_{7} a_{2} \dots a_{5}}}_{0} \Big) \\
= &\frac{11!}{4(4!)^2} \; \Big( 2 \; \frac{1}{11} \; R_{\alpha_1 \alpha_2 c_1 c_2} \; {R_{\alpha_3 a_1}}^{c_2 c_1} \; G_{a_2 a_3 \alpha_4 \alpha_5} \; G_{a_4 a_5 \alpha_6 \alpha_7} \Big),
\end{align*}
which becomes, using (\ref{eq_fieldcomp}),
\begin{align}
&(\gamma^{f})_{\alpha_1 \alpha_2} \; K_{f a_{1} \dots a_{5} \alpha_{3} \dots \alpha_7} = \notag \\
& -120 \;i\; (\gamma_{a_1 \dots a_5})_{\alpha_1 \alpha_2} {R_{\alpha_3 \alpha_4}}^{c_1 c_2} \; \bigg( (\gamma^{e_1 e_2})_{\alpha_5 \alpha_6} (\gamma_{[c_1 c_2} \; {T_{e_1 e_2]}})_{\alpha_7} + \frac{1}{24} ({\gamma_{c_1 c_2}}^{e_1 \dots e_4})_{\alpha_5 \alpha_6} (\gamma_{e_1 e_2} \; {T_{e_3 e_4}})_{\alpha_7} \bigg) \notag \\
& + 1800 \;i\; (\gamma_{a_1 a_2})_{\alpha_1 \alpha_2} \; (\gamma_{a_3 a_4})_{\alpha_3 \alpha_4} {R_{\alpha_5 \alpha_6}}^{c_1 c_2} R_{\alpha_7 a_5 c_2 c_1}~.
\label{eq_Kd52L3}
\end{align}
The second term in (\ref{eq_Kd52L3}) can be written,
\begin{align}
& 1800 \vphantom{\frac{1}{1}} \;i\; (\gamma_{[a_1 a_2|})_{\alpha_1 \alpha_2} \; (\gamma_{|a_3 a_4|})_{\alpha_3 \alpha_4} \; {R_{\alpha_5 \alpha_6}}^{c_1 c_2} R_{\alpha_7 |a_5] c_2 c_1} \label{eq_Kd52L3_2} \\
 \ &= 600 \;i\; (\gamma^{g})_{\alpha_1 \alpha_2} \; (\gamma_{g [a_1 a_2 a_3 a_4|})_{\alpha_3 \alpha_4} \; {R_{\alpha_5 \alpha_6}}^{c_1 c_2} R_{\alpha_7 |a_5] c_2 c_1} \notag \\
\ &= 600 \;i\; (\gamma^{g})_{\alpha_1 \alpha_2} \bigg( \frac{6}{5} (\gamma_{[g a_1 a_2 a_3 a_4|})_{\alpha_3 \alpha_4} \; {R_{\alpha_5 \alpha_6}}^{c_1 c_2} R_{\alpha_7 |a_5] c_2 c_1} \;+\; \frac{1}{5} (\gamma_{a_1 a_2 a_3 a_4 a_5})_{\alpha_3 \alpha_4} \; {R_{\alpha_5 \alpha_6}}^{c_1 c_2} R_{\alpha_7 gc_2 c_1} \bigg) \notag~.
\end{align}
One can then verify that the second term on the right-hand side of (\ref{eq_Kd52L3_2}) cancels with the  first term on the right-hand side of (\ref{eq_Kd52L3}). Since the first term on the right-hand side of (\ref{eq_Kd52L3_2}) is in a $\tau$-exact form and 
$H_{\tau}^{6,5}$ is trivial, the solution reads,
\begin{align*}
K^{(5/2)}_{a_1 \dots a_6 \alpha_1 \dots \alpha_5} = 720 \;i\; (\gamma_{a_1 \dots a_5})_{\alpha_1 \alpha_2} \; {R_{\alpha_3 \alpha_4}}^{c_1 c_2} R_{\alpha_5 a_6 c_2 c_1}~,
\end{align*}
up to $\tau$-exact terms.

\subsection*{Dimension 3 - ($A_{1} \dots A_{6} \rightarrow \alpha_1 \dots \alpha_{6}$, $A_{7} \dots A_{12} \rightarrow a_1 \dots a_{6}$)}

At dimension $3$, eq.~(\ref{eq_dKeqGcompL3}) reads,
\begin{align*}
& \frac{1}{2} \; D_{\alpha_{1}} K_{\alpha_{2} \dots \alpha_{6} a_{1} \dots a_{6}} \;+\; \frac{1}{2} \; D_{a_{1}} \overbrace{K_{a_{2} \dots a_{6} \alpha_{1} \dots \alpha_{6}}}^{0} \\
+ &\frac{11}{2} \left( \frac{5}{22} \; {T_{\alpha_{1} \alpha_{2}}}^{f} \; K_{f a_{1} \dots a_{6} \alpha_{3} \dots \alpha_6} \;+\; \frac{5}{22} \; {T_{a_{1} a_{2}}}^{\gamma} \; \underbrace{K_{\gamma \alpha_{1} \dots \alpha_{6} a_{3} \dots a_{6}}}_{0} \;+\; \frac{6}{11} \; {T_{a_{1} \alpha_{1}}}^{\gamma} \; \underbrace{K_{\gamma \alpha_{3} \dots \alpha_{6} a_{2} \dots a_{6}}}_{0} \right) \\
= &\frac{11!}{4(4!)^2} \bigg( -\frac{12}{77} \; {R_{\alpha_1 a_1}}^{c_1 c_2} \; R_{\alpha_2 a_2 c_2 c_1} \; G_{a_3 a_4 \alpha_3 \alpha_4} \; G_{a_5 a_6 \alpha_5 \alpha_6} \\
& \qquad \qquad \qquad \qquad \qquad + 2 \; \frac{1}{154} \; {R_{\alpha_1 \alpha_2}}^{c_1 c_2} \; R_{\alpha_3 \alpha_4 c_2 c_1} \; G_{a_1 a_2 \alpha_5 \alpha_6} \; G_{a_3 \dots a_6} \\
& \qquad \qquad \qquad \qquad \qquad \qquad \qquad \qquad + 2 \; \frac{3}{77} \; {R_{a_1 a_2}}^{c_1 c_2} \; R_{\alpha_1 \alpha_2 c_2 c_1} \; G_{a_3 a_4 \alpha_3 \alpha_4} \; G_{a_5 a_6 \alpha_5 \alpha_6} \bigg)~,
\end{align*}
which becomes, using (\ref{eq_fieldcomp}),
\begin{align*}
- \frac{5}{4} \;i\; (\gamma^{f})_{\alpha_1 \alpha_2} \; K_{f a_1 \dots a_6 \alpha_3 \dots \alpha_6} = &-\frac{1}{2} \; D_{\alpha_1} \; K_{\alpha_2 \dots \alpha_6 a_1 \dots a_6} \notag \\
&- \frac{1}{2} \; \d_{a_1} \; K_{a_2 \dots a_6 \alpha_1 \dots \alpha_6} \notag \\
&- 3 \; T_{a_1 \alpha_1}^{\epsilon} \; K_{\epsilon \alpha_2 \dots \alpha_6 a_2 \dots a_6} \notag \\
&- 2700 \; {R_{\alpha_1 a_1}}^{c_1 c_2} \; R_{\alpha_2 a_2 c_2 c_1} \; G_{a_3 a_4 \alpha_3 \alpha_4} \; G_{a_5 a_6 \alpha_5 \alpha_6} \notag \\
&+ 225 \; {R_{\alpha_1 \alpha_2}}^{c_1 c_2} \; R_{\alpha_3 \alpha_4 c_2 c_1} \; G_{a_1 a_2 \alpha_5 \alpha_6} \; G_{a_3 \dots a_6} \notag \\
&+ 1350 \; {R_{a_1 a_2}}^{c_1 c_2} \; R_{\alpha_1 \alpha_2 c_2 c_1} \; G_{a_3 a_4 \alpha_3 \alpha_4} \; G_{a_5 a_6 \alpha_5 \alpha_6} \notag~.
\end{align*}
Let us now examine separately each group of terms in the equation above  with the same type of field content. 
There are four $G^3$ terms which read,
\begin{align}
& -225 \;i\; G_{a_1 \dots a_4} \; (\gamma_{a_5 a_6})_{\alpha_1 \alpha_2} \; {R_{\alpha_3 \alpha_4}}^{c_1 c_2} \; R_{\alpha_5 \alpha_6 c_1 c_2} \notag \\
& -360 \; (\gamma_{a_1 \dots a_5})_{\alpha_1 \alpha_2} \; R_{\alpha_3 \alpha_4 c_1 c_2} \; {T_{c_2 \alpha_5}}^{\epsilon} \; {T_{c_1 \epsilon}}^{\beta} \; (\gamma_{a_6})_{\beta \alpha_6} \notag \\
& +720 \; (\gamma_{a_1 \dots a_5})_{\alpha_1 \alpha_2} \; R_{\alpha_3 \alpha_4 c_1 c_2} \; {T_{c_1 \alpha_5}}^{\epsilon} \; {T_{a_6 \epsilon}}^{\beta} \; (\gamma_{c_2})_{\beta \alpha_6} \notag \\
& -540 \;i\; (\gamma_{a_1 \dots a_5})_{(\alpha_1 \epsilon|} \; {T_{a_6 \alpha_2}}^{\epsilon} \; {R_{|\alpha_3 \alpha_4|}}^{c_1 c_2} \; R_{|\alpha_5 \alpha_6) c_2 c_1}~.
\label{eq_dK3_GGG}
\end{align}
The last term in (\ref{eq_dK3_GGG}) can be split in two parts,
\begin{align*}
-540 \;i\; \bigg( \frac{2}{6}(\gamma_{a_1 \dots a_5})_{\alpha_1 \epsilon} \; {T_{a_6 \alpha_2}}^{\epsilon} \; {R_{\alpha_3 \alpha_4}}^{c_1 c_2} \; R_{\alpha_5 \alpha_6 c_2 c_1} + \frac{4}{6}(\gamma_{a_1 \dots a_5})_{\alpha_1 \alpha_2} \; {T_{a_6 \alpha_3}}^{\epsilon} \; {R_{\epsilon \alpha_4}}^{c_1 c_2} \; R_{\alpha_5 \alpha_6 c_2 c_1} \bigg)~.
\end{align*}
The first one leads to,
\begin{align*}
(\gamma^{g})_{\alpha_1 \alpha_2} \bigg( \frac{5}{8} \;i\; {\epsilon_{g a_1 \dots a_6}}^{b_1 \dots b_4} \; G_{b_1 \dots b_4} \; {R_{\alpha_3 \alpha_4}}^{c_1 c_2} \; R_{\alpha_5 \alpha_6 c_2 c_1} \bigg) \;+\; 225 \;i\; G_{a_1 \dots a_4} \; (\gamma_{a_5 a_6})_{\alpha_1 \alpha_2} \; {R_{\alpha_3 \alpha_4}}^{c_1 c_2} \; R_{\alpha_5 \alpha_6 c_2 c_1}~,
\end{align*}
where the first term is $\tau$-exact, and the second term cancels with the first one in (\ref{eq_dK3_GGG}). It can then be verified that the three remaining $G^3$ terms cancel out. 

Moreover there are three terms of the schematic form $G (DG)$,
\begin{align}
&-360 \;i\; (\gamma^{a_1 \dots a_5})_{\alpha_1 \alpha_2} \; {R_{\alpha_3 \alpha_4}}^{c_1 c_2} \; \d_{c_2} {T_{c_1 \alpha_5}}^{\beta} (\gamma_{a_6})_{\beta \alpha_6} \notag \\
&- 720 \;i\; (\gamma^{a_1 \dots a_5})_{\alpha_1 \alpha_2} \; {R_{\alpha_3 \alpha_4}}^{c_1 c_2} \; \d_{c_1} {T_{a_6 \alpha_5}}^{\beta} (\gamma_{c_2})_{\beta \alpha_6} \notag \\
&-180 \;i\; (\gamma^{a_1 \dots a_5})_{\alpha_1 \alpha_2} \; {R_{\alpha_3 \alpha_4}}^{c_1 c_2} \; \d_{a_6} R_{\alpha_5 \alpha_6 c_2 c_1}~,
\label{eq_dK3_GdG}
\end{align}
which cancel out.

There are two $RG$ terms which read,
\begin{align}
&-1350 \; (\gamma_{a_1 a_2})_{\alpha_1 \alpha_2} \; (\gamma_{a_3 a_4})_{\alpha_3 \alpha_4} \; {R_{a_5 a_6}}^{c_1 c_2} \; R_{\alpha_5 \alpha_6 c_2 c_1} \notag \\
&+45 \; (\gamma_{a_1 \dots a_5})_{\alpha_1 \alpha_2} \; {R_{\alpha_3 \alpha_4}}^{c_1 c_2} \; \Big( (\gamma^{e_1 e_2} \; \gamma_{a_6})_{\alpha_5 \alpha_6} \; R_{c_2 c_1 e_1 e_2} - 2 \; (\gamma^{e_1 e_2} \; \gamma_{c_2})_{\alpha_5 \alpha_6} \; R_{c_1 a_6 e_1 e_2} \Big)~.
\label{eq_dK3_RG}
\end{align}
The first term of (\ref{eq_dK3_RG})can be put in a $\tau$-exact form,
\begin{align*}
&-1350 \; (\gamma_{a_1 a_2})_{\alpha_1 \alpha_2} \; (\gamma_{a_3 a_4})_{\alpha_3 \alpha_4} \; {R_{a_5 a_6}}^{c_1 c_2} \; R_{\alpha_5 \alpha_6 c_2 c_1} =\\
& \qquad (\gamma^{g})_{\alpha_1 \alpha_2} \Big( -630 \;i\; (\gamma_{[g a_1 \dots a_4|})_{\alpha_3 \alpha_4} \; {R_{|a_5 a_6]}}^{c_1 c_2} \; R_{\alpha_5 \alpha_6 c_2 c_1} + 180 \; (\gamma_{a_1 \dots a_5})_{\alpha_1 \alpha_2} \; {R_{a_6 g}}^{c_1 c_2} \; R_{\alpha_5 \alpha_6 c_2 c_1} \Big)~,
\end{align*}
while the remaining $RG$ terms cancel out. 

There are two $T^2$ terms which read,
\begin{align}
&+2700 \; (\gamma_{a_1 a_2})_{\alpha_1 \alpha_2} \; (\gamma_{a_3 a_4})_{\alpha_3 \alpha_4} \;  {R_{\alpha_5 a_5}}^{c_1 c_2} \; R_{\alpha_6 a_6 c_2 c_1} \label{eq_dK3_TT} \\
&+1080 \;i\; (\gamma_{a_1 \dots a_5})_{\alpha_1 \alpha_2} \; \Big( (\gamma^{e_1 e_2})_{\alpha_3 \alpha_4} (\gamma_{[c_1 c_2} \; {T_{e_1 e_2]}})_{\alpha_5} + \frac{1}{24} ({\gamma_{c_1 c_2}}^{e_1 \dots e_4})_{\alpha_3 \alpha_4} (\gamma_{e_1 e_2} \; {T_{e_3 e_4}})_{\alpha_5} \Big) R_{\alpha_6 a_6 c_2 c_1} \notag~.
\end{align}
The first term can be put in a $\tau$-exact form,
\begin{align*}
&2700 \; (\gamma_{a_1 a_2})_{\alpha_1 \alpha_2} \; (\gamma_{a_3 a_4})_{\alpha_3 \alpha_4} \;  {R_{\alpha_5 a_5}}^{c_1 c_2} \; R_{\alpha_6 a_6 c_2 c_1} = \notag \\
& \qquad \frac{1}{3} \; 2700 \; (\gamma^{g})_{\alpha_1 \alpha_2} \bigg( \frac{7}{5} \; (\gamma_{g a_1 \dots a_4})_{\alpha_3 \alpha_4} \; {R_{\alpha_5 a_5}}^{c_1 c_2} \; R_{\alpha_6 a_6 c_2 c_1} + \frac{2}{5} \; (\gamma_{a_1 \dots a_5})_{\alpha_3 \alpha_4} \; {R_{\alpha_5 a_6}}^{c_1 c_2} \; R_{\alpha_6 g c_2 c_1} \bigg)~,
\end{align*}
while the remaining $TT$ terms cancel out. Taking the triviality of $H_{\tau}^{7,4}$ into account, 
the non-vanishing terms extracted from the $RG$, $T^2$, and $G^3$ terms lead to the solution,
\begin{align}
K^{(3)}_{a_1 \dots a_7 \alpha_1 \dots \alpha_4} =& \; 504 \;i\; (\gamma_{a_1 \dots a_5})_{\alpha_1 \alpha_2} \Big( - {R_{a_6 a_7}}^{c_1 c_2} \; R_{\alpha_5 \alpha_6 c_2 c_1} + 2 \; {R_{\alpha_5 a_6}}^{c_1 c_2} \; R_{\alpha_6 a_7 c_2 c_1} \Big) \notag \\
&-\; \frac{1}{2} \;i\; {\epsilon_{a_1 \dots a_7}}^{b_1 \dots b_4} \; G_{b_1 \dots b_4} \; {R_{\alpha_1 \alpha_2}}^{c_1 c_2} \; R_{\alpha_3 \alpha_4 c_2 c_1}~,
\end{align}
up to $\tau$-exact terms.

\subsection*{Dimensions 7/2 - ($A_{1} \dots A_{5} \rightarrow \alpha_1 \dots \alpha_{5}$, $A_{6} \dots A_{12} \rightarrow a_1 \dots a_{7}$)}

At dimension $7/2$, eq.~(\ref{eq_dKeqGcompL3}) reads,
\begin{align*}
& \frac{5}{12} \; D_{\alpha_{1}} K_{\alpha_{2} \dots \alpha_{5} a_{1} \dots a_{7}} \;-\; \frac{7}{12} \; D_{a_{1}} K_{a_{2} \dots a_{7} \alpha_{1} \dots \alpha_{5}} \\
- &\frac{11}{2} \left( \frac{5}{33} \; {T_{\alpha_{1} \alpha_{2}}}^{f} \; K_{f a_{1} \dots a_{7} \alpha_{3} \dots \alpha_5} \;-\; \frac{7}{22} \; {T_{a_{1} a_{2}}}^{\gamma} \; K_{\gamma \alpha_{1} \dots \alpha_{5} a_{3} \dots a_{7}} \;+\; \frac{35}{66} \; {T_{a_{1} \alpha_{1}}}^{\gamma} \; K_{\gamma \alpha_{3} \dots \alpha_{5} a_{2} \dots a_{7}} \right) \\
&= \frac{11!}{4(4!)^2} \bigg( 2 \; \frac{1}{11} \; {R_{\alpha_1 a_1}}^{c_1 c_2} \; R_{a_2 a_3 c_2 c_1} \; G_{a_4 a_5 \alpha_3 \alpha_4} \; G_{a_6 a_7 \alpha_5 \alpha_6} \\
& \qquad \qquad \qquad \qquad \qquad \qquad \qquad \qquad \qquad + 4 \; \frac{1}{66} \; {R_{\alpha_1 a_1}}^{c_1 c_2} \; R_{\alpha_2 \alpha_3 c_2 c_1} \; G_{a_2 a_3 \alpha_4 \alpha_5} \; G_{a_4 \dots a_7} \bigg)~.
\end{align*}
The right-hand side of the equation above contains terms of the form $G (DT)$, $T (DG)$, $TR$, and $TG^2$. The first two groups of terms simply vanish (without the use of any equations of motion or BI). Two $\tau$-exact terms can be extracted from $RT$ and $TG^2$, and the remaining terms cancel out. This leads to the solution,
\begin{align*}
K^{(7/2)}_{a_1 \dots a_8 \alpha_1 \dots \alpha_3} =& \;2016 \;i\; (\gamma_{a_1 \dots a_5})_{\alpha_1 \alpha_2} \; {R_{a_6 a_7}}^{c_1 c_2} R_{\alpha_3 a_8 c_2 c_1} \\
&+ \; 4 \; {\epsilon_{a_1 \dots a_7}}^{b_1 \dots b_4} \; G_{b_1 \dots b_4} \; {R_{\alpha_1 \alpha_2}}^{c_1 c_2} \; R_{\alpha_3 a_8 c_2 c_1}~,
\end{align*}
up to $\tau$-exact terms.

\subsection*{Dimensions 4 - ($A_{1} \dots A_{4} \rightarrow \alpha_1 \dots \alpha_{4}$, $A_{5} \dots A_{12} \rightarrow a_1 \dots a_{8}$)}

At dimension $4$, eq.~(\ref{eq_dKeqGcompL3}) reads,
\begin{align*}
& \frac{4}{12} \; D_{\alpha_{1}} K_{\alpha_{2} \dots \alpha_{4} a_{1} \dots a_{8}} \;+\; \frac{8}{12} \; D_{a_{1}} K_{a_{2} \dots a_{8} \alpha_{1} \dots \alpha_{4}} \\
+ &\frac{11}{2} \left( \frac{1}{11} \; {T_{\alpha_{1} \alpha_{2}}}^{f} \; K_{f a_{1} \dots a_{8} \alpha_3 \alpha_4} \;+\; \frac{14}{33} \; {T_{a_{1} a_{2}}}^{\gamma} \; K_{\gamma \alpha_{1} \dots \alpha_{4} a_{3} \dots a_{8}} \;+\; \frac{16}{33} \; {T_{a_{1} \alpha_{1}}}^{\gamma} \; K_{\gamma \alpha_{2} \dots \alpha_{4} a_{2} \dots a_{8}} \right) \\
= &\frac{11!}{4(4!)^2} \bigg( 1260 \; {R_{a_1 a_2}}^{c_1 c_2} \; R_{a_3 a_4 c_2 c_1} \; G_{a_5 a_6 \alpha_3 \alpha_4} \; G_{a_7 a_8 \alpha_5 \alpha_6} \\
& \qquad \qquad \qquad \qquad + 35 \; {R_{\alpha_1 \alpha_2}}^{c_1 c_2} \; R_{\alpha_3 \alpha_4 c_2 c_1} \; G_{a_1 a_2 a_3 a_4} \; G_{a_5 \dots a_8} \\
& \qquad \qquad \qquad \qquad \qquad \qquad + 4 \cdot 210 \; {R_{\alpha_1 \alpha_2}}^{c_1 c_2} \; R_{a_1 a_2 c_2 c_1} \; G_{a_3 a_4 a_5 a_6} \; G_{a_7 a_8 \alpha_3 \alpha_4} \\
& \qquad \qquad \qquad \qquad \qquad \qquad \qquad \qquad - 2 \cdot 840 \; {R_{\alpha_1 a_1}}^{c_1 c_2} \; R_{\alpha_2 a_2 c_2 c_1} \; G_{a_3 a_4 \alpha_3 \alpha_4} \; G_{a_5 a_6 a_7 a_8} \bigg)~.
\end{align*}
The terms in the equation above can be cast in eight groups: $R^2$, $RG^2$, $R (DG)$, $G^4$, $G^2 (DG)$, $G T^2$, $T (DT)$, and $G (DR)$. 
Parts of the terms of the form $R^2$, $G^2 R$, and $G T^2$ can be put in a $\tau$-exact form, while the remaining terms cancel out. Taking into account the BI,
\begin{align}
D_{a_1} R_{a_2 a_3 c_1c_2} = - {T_{a_1 a_2}}^{\gamma} R_{\gamma a_3 c_1c_2}~,
\label{eq_BI_dRTT}
\end{align}
we see that the term $G (DR)$ cancel against a term from $GT^2$. Taking into account the equation of motion of $G$ we see that a term from $G^2 (DG)$  cancels against a term in $G^4$,
\begin{align}
{\epsilon_{a_1 \dots a_7}}^{b_1 \dots b_4} \; D_{a_8} G_{b_1 \dots b_4} = \frac{1}{2} \; {\epsilon_{a_1 \dots a_8}}^{b_1 \dots b_3} \; D^{c} G_{c b_1 \dots b_3} = 105 \; G_{a_1 \dots a_4} \; G_{a_5 \dots a_8}~.
\label{eq_EOMG}
\end{align}
We are thus led to the solution,
\begin{align*}
K^{(4)}_{a_1 \dots a_9 \alpha_1 \dots \alpha_2} = \;-&1512 \;i\; (\gamma_{a_1 \dots a_5})_{\alpha_1 \alpha_2} \; {R_{a_6 a_7}}^{c_1 c_2} \; R_{a_8 a_9 c_2 c_1} \\
-& \; 6 \; {\epsilon_{a_1 \dots a_7}}^{b_1 \dots b_4} \; G_{b_1 \dots b_4} \; {R_{\alpha_1 \alpha_2}}^{c_1 c_2} \; R_{a_8 a_9 c_2 c_1} \\
+& \; 12 \; {\epsilon_{a_1 \dots a_7}}^{b_1 \dots b_4} \; G_{b_1 \dots b_4} \; {R_{\alpha_1 a_8}}^{c_1 c_2} \; R_{\alpha_2 a_9 c_2 c_1}~,
\end{align*}
up to $\tau$-exact terms.

\subsection*{Dimensions 9/2 - ($A_{1} \dots A_{3} \rightarrow \alpha_1 \dots \alpha_{3}$, $A_{4} \dots A_{12} \rightarrow a_1 \dots a_{9}$)}

At dimension $9/2$, eq.~(\ref{eq_dKeqGcompL3}) reads,
\begin{align*}
& \frac{3}{12} \; D_{\alpha_{1}} K_{\alpha_{2} \alpha_{3} a_{1} \dots a_{9}} \;-\; \frac{9}{12} \; D_{a_{1}} K_{a_{2} \dots a_{9} \alpha_{1} \dots \alpha_{3}} \\
- &\frac{11}{2} \left( \frac{1}{22} \; {T_{\alpha_{1} \alpha_{2}}}^{f} \; K_{f a_{1} \dots a_{9} \alpha_3} \;+\; \frac{6}{11} \; {T_{a_{1} a_{2}}}^{\gamma} \; K_{\gamma \alpha_{1} \dots \alpha_{3} a_{3} \dots a_{9}} \;+\; \frac{9}{22} \; {T_{a_{1} \alpha_{1}}}^{\gamma} \; K_{\gamma \alpha_{2} \alpha_{3} a_{2} \dots a_{9}} \right) \\
= &\frac{11!}{4(4!)^2} \bigg( 2 \; \frac{1}{110} \; {R_{\alpha_1 \alpha_2}}^{c_1 c_2} \; R_{\alpha_3 a_1 c_2 c_1} \; G_{a_2 \dots a_5} \; G_{a_6 \dots a_9} \\
& \qquad \qquad \qquad \qquad \qquad \qquad \qquad \qquad + 4 \; \frac{3}{55} \; {R_{a_1 a_2}}^{c_1 c_2} \; R_{\alpha_1 a_3 c_2 c_1} \; G_{\alpha_2 \alpha_3 a_4 a_5} \; G_{a_6 \dots a_9} \bigg)~.
\end{align*}
The terms  in the equation above can be cast in seven groups: $R (DT)$, $RTG$, $G^2(DT)$, $G^3 T$, $T^3$, $TG(DG)$ and $T(DR)$. 
One term of the form $RTG$ is $\tau$-exact, while all the remaining terms can be seen to cancel out, using (\ref{eq_BI_dRTT}) and (\ref{eq_EOMG}) to convert a term of the form $T (DR)$ to  the form $T^3$, and a term of the form $TG(DG)$ to the form $G^3 T$. 
Up to $\tau$-exact terms, the component of dimension $9/2$ then reads, $$ K^{(9/2)}_{a_1 \dots a_9 \alpha_1 \alpha_2} = \; 60 \;i\; {\epsilon_{a_1 \dots a_7}}^{b_1 \dots b_4} \; G_{b_1 \dots b_4} \; {R_{a_8 a_9}}^{c_1 c_2} \; R_{\alpha_1 a_{10} c_2 c_1}~. $$

\subsection*{Dimensions 5 - ($A_{1} A_{2} \rightarrow \alpha_1 \alpha_{2}$, $A_{3} \dots A_{12} \rightarrow a_1 \dots a_{10}$)}

At dimension $5$, eq.~(\ref{eq_dKeqGcompL3}) reads,
\begin{align*}
& \frac{2}{12} \; D_{\alpha_{1}} K_{\alpha_{2} a_{1} \dots a_{9}} \;+\; \frac{10}{12} \; D_{a_{1}} K_{a_{2} \dots a_{10} \alpha_{1} \alpha_{2}} \\
+ &\frac{11}{2} \left( \frac{1}{66} \; {T_{\alpha_{1} \alpha_{2}}}^{f} \; K_{f a_{1} \dots a_{10}} \;+\; \frac{15}{22} \; {T_{a_{1} a_{2}}}^{\gamma} \; K_{\gamma \alpha_{1} \alpha_{2} a_{3} \dots a_{10}} \;+\; \frac{10}{33} \; {T_{a_{1} \alpha_{1}}}^{\gamma} \; K_{\gamma \alpha_{2} a_{2} \dots a_{10}} \right) \\
= &\frac{11!}{4(4!)^2} \bigg( 2 \; \frac{1}{66} \; {R_{\alpha_1 \alpha_2}}^{c_1 c_2} \; R_{a_1 a_2 c_2 c_1} \; G_{a_3 \dots a_6} \; G_{a_7 \dots a_{10}} \\
& \qquad \qquad \qquad \qquad \quad - 1 \; \frac{2}{33} \; {R_{\alpha_1 a_1}}^{c_1 c_2} \; R_{\alpha_2 a_2 c_2 c_1} \; G_{a_3 a_4 a_5 a_6} \; G_{a_7 \dots a_{10}} \\
& \qquad \qquad \qquad \qquad \qquad \qquad \qquad + 2 \; \frac{1}{11} \; {R_{a_1 a_2}}^{c_1 c_2} \; R_{a_3 a_4 c_2 c_1} \; G_{a_5 a_6 \alpha_1 \alpha_2} \; G_{a_7 a_8 \alpha_3 \alpha_4} \bigg)~.
\end{align*}
The terms in the equation above can be cast in nine groups: $RT^2$, $GT(DT)$, $G^2 T^2$, $G R^2$, $GR(DG)$, $R G^3$, $R(DR)$, $G^2(DR)$, and $T^2(DG)$. 
One term in $G R^2$ is $\tau$-exact, while all the remaining terms cancel out, as can be seen using eq. (\ref{eq_BI_dRTT}) and (\ref{eq_EOMG}) to convert a term of the form 
 $R(DR)$ to the form $RT^2$, a term of the form $G^2(DR)$ to the form $G^2 T^2$, and a term of the form  $T^2 (DG)$ to the form $G^2 T^2$. 
Up to $\tau$-exact terms, the component of dimension $5$ then reads,  
\eq{\label{topl3}
K^{(5)}_{a_1 \dots a_{11}} = \; -165 \; {\epsilon_{a_1 \dots a_7}}^{b_1 \dots b_4} \; G_{b_1 \dots b_4} \; {R_{a_8 a_9}}^{c_1 c_2} \; R_{a_{10} a_{11} c_2 c_1} ~.
}

\subsection*{Dimensions 11/2 - ($A_{1} \rightarrow \alpha_1 $, $A_{2} \dots A_{12} \rightarrow a_1 \dots a_{11}$)}

Since there is no new component of $K$ appearing, this equation should be satisfied automatically,
\begin{align*}
& \frac{1}{12} \; D_{\alpha_{1}} K_{a_{1} \dots a_{11}} -\; \frac{11}{12} \; D_{a_{1}} K_{a_{2} \dots a_{11} \alpha_1} \;- \frac{11}{2} \left( \frac{1}{6} \; {T_{a_1 \alpha_{2}}}^{f} \; K_{f a_{2} \dots a_{11}} -\; \frac{5}{6} \; {T_{a_{1} a_{2}}}^{\gamma} \; K_{\gamma \alpha_{1} a_{3} \dots a_{11}} \right) \\
&= \frac{11!}{4(4!)^2} \bigg( \frac{2}{6} \; {R_{\alpha_1 a_1}}^{c_1 c_2} \; R_{a_2 a_3 c_2 c_1} \; G_{a_4 \dots a_7} \; G_{a_8 \dots a_{11}} \bigg)~.
\end{align*}
The equation contains six types of terms: $T R^2$, $GR(DT)$, $G^2 TR$, $GT(DR)$, $RT(DG)$, and $T^3 G$. As expected all the terms cancel out, as can be seen using  (\ref{eq_BI_dRTT}) and (\ref{eq_EOMG}) to convert a term of the form $GT(DR)$ to the form $T^3 G$, and a term of the form $RT(DG)$ to the form $G^2 TR$.

\subsection*{Action at $\mathcal{O}(l^3)$}

We have  constructed the explicit expression of each component of $K_{11}$ and showed that it is unique up to exact terms. 
In particular the top component, given in eq.~(\ref{topl3}), precisely agrees with (\ref{tyf}), leading to the superinvariant of section \ref{sec:five}.


\section{Weil triviality at $l^6$}\label{app:6}

The same method will be used to generate the corrections at $l^6$-order, cf. section \ref{sec:eight}. We will look for the solution to the equation  
$\d K_{11} = G_4 \swed X^{(1)}_8 $. In components this reads,
\begin{align}
& D_{[A_{1}} \; K_{A_{2} \dots A_{12})} + \frac{11}{2} \; T_{[A_{1} A_{2}|}^{\qquad F} \; K_{F|A_{3} \dots A_{12})} = \notag \\
& \qquad \qquad \frac{11!}{(4!) 4^2} \bigg(  G_{[A_{1} \dots A_{4}} \; R_{|A_{5} A_{6}|c_1 c_2} \; {R_{|A_{7} A_{8}|}}^{c_2 c_3} \; R_{|A_{9} A_{10}|c_3 c_4} \; {R_{|A_{11} A_{12})}}^{c_4 c_1} \notag \\
& \qquad \qquad \qquad \qquad \qquad - \frac{1}{4} \; G_{[A_{1} \dots A_{4}} \; R_{|A_{5} A_{6}|c_1 c_2} \; {R_{|A_{7} A_{8}|}}^{c_1 c_2} \; R_{|A_{9} A_{10}|d_1 d_2} \; {R_{|A_{11} A_{12})}}^{d_1 d_2} \bigg)~.
\label{eq_dKeqGcompL6}
\end{align}
The dimensions of the various components of $K_{11}$ now range from $\left[ K_{\alpha_1 \dots \alpha_{12}} \right] = \frac{5}{2}$ to $\left[ K_{a_1 \dots a_{11}} \right] = 8$.

\subsection*{Dimension 3 and 7/2}

If we assume that the superivariant at $\mathcal{O}(l^6)$ is quartic or higher in fields, the first potentially nonvanishing component of $K_{11}$  appears at dimension 4 (it is of the form  $G^4$).  We thus obtain,
\eq{
K^{(5/2)}_{\alpha_1 \dots \alpha_{11}}= K^{(3)}_{a_1 \alpha_1 \dots \alpha_{10}}= K^{(7/2)}_{a_1 a_2 \alpha_1 \dots \alpha_9}=0~.
\nonumber}
This is consistent with (\ref{eq_dKeqGcompL6}), whose right-hand side vanishes for dimensions lower than 4.

\subsection*{Dimension 4 - ($ A_{1} \dots A_{10} \rightarrow \alpha_1 \dots \alpha_{10}$, $A_{11} A_{12} \rightarrow a_1 a_2 $)}

Eq.~(\ref{eq_dKeqGcompL6}) takes the following form,
\begin{align*}
& \frac{2}{12} \; D_{a_{1}} \overbrace{K^{(3)}_{a_{2} \alpha_{1} \dots \alpha_{10}}}^{0} + \frac{10}{12} \; D_{\alpha_{1}} \overbrace{K^{(7/2)}_{\alpha_2 \dots \alpha_{10} a_1 a_2}}^{0} \\
+ & \; \frac{11}{2} \bigg( \frac{15}{22} \; {T_{\alpha_1 \alpha_2}}^{f} \;K^{(4)}_{f a_1 a_2 \alpha_3 \dots \alpha_{10}} + \frac{10}{33} \; {T_{a_1 \alpha_1}}^{\gamma} \; \overbrace{K^{(3)}_{\gamma \alpha_2 \dots \alpha_{10} a_2}}^{0} + \frac{1}{66} \; {T_{a_1 a_2}}^{\gamma} \; \overbrace{K^{(5/2)}_{\gamma \alpha_1 \dots \alpha_{10}}}^{0} \bigg) \\
= & \; \frac{11!}{4! 4^2} \; G_{a_1 a_2 \alpha_1 \alpha_2} \bigg( {R_{\alpha_3 \alpha_4}}^{c_1 c_2} \; {R_{\alpha_5 \alpha_6}}_{c_2 c_3} \; {R_{\alpha_7 \alpha_8}}^{c_3 c_4} \; {R_{\alpha_9 \alpha_{10}}}_{c_4 c_1} \\
& \qquad \qquad \qquad \qquad \qquad \qquad \qquad - \frac{1}{4} \; {R_{\alpha_3 \alpha_4}}^{c_1 c_2} \; {R_{\alpha_5 \alpha_6}}_{c_1 c_2} \; {R_{\alpha_7 \alpha_8}}^{d_1 d_2} \; {R_{\alpha_9 \alpha_{10}}}_{d_1 d_2} \bigg)~,
\end{align*}
which simplifies to,
\begin{align}
& (\gamma^f)_{\alpha_1 \alpha_2} \; K^{(4)}_{f a_1 a_2 \alpha_3 \dots \alpha_{10}}= 2520 \; (\gamma_{a_1 a_2})_{\alpha_1 \alpha_2} \bigg( {R_{\alpha_3 \alpha_4}}^{c_1 c_2} \; {R_{\alpha_5 \alpha_6}}_{c_2 c_3} \; {R_{\alpha_7 \alpha_8}}^{c_3 c_4} \; {R_{\alpha_9 \alpha_{10}}}_{c_4 c_1} \notag\\
& \qquad \qquad \qquad \qquad \qquad \qquad \qquad \qquad \qquad - \frac{1}{4} \; {R_{\alpha_3 \alpha_4}}^{c_1 c_2} \; {R_{\alpha_5 \alpha_6}}_{c_1 c_2} \; {R_{\alpha_7 \alpha_8}}^{d_1 d_2} \; {R_{\alpha_9 \alpha_{10}}}_{d_1 d_2} \bigg) \notag \\
& \qquad \qquad = (\gamma_{a_1 a_2})_{\alpha_1 \alpha_2} \; X^{(8)}_{\alpha_3 \dots \alpha_{10}}.
\label{eq_Kd4L3}
\end{align}
%


Explicitly, the term $(\mathrm{tr}R^2)^2$ reads (omitting the factor $- 1/4$),
\begin{align*}
&\tfrac{1}{6^4} \; (\gamma^{u_0 u_1})(\gamma^{u_2 u_3})(\gamma^{u_4 u_5})(\gamma^{u_6 u_7}) \;\; {G_{u_0 u_1}}^{y_0 y_1} G_{u_2 u_3 y_0 y_1} G_{u_4 u_5 z_0 z_1} {G_{u_6 u_7}}^{z_0 z_1} \\
&\tfrac{4}{24 \cdot 6^4} \; (\gamma^{u_0 u_1})(\gamma^{u_2 u_3})(\gamma^{u_4 u_5})(\gamma^{v_0 \dots v_5}) \;\; {G_{u_0 u_1}}^{y_0 y_1} G_{u_2 u_3 y_0 y_1} G_{u_4 u_5 v_0 v_1} G_{v_2 \dots v_5} \\
&\tfrac{2}{24^2 \; 6^4} \; (\gamma^{u_0 u_1})(\gamma^{u_2 u_3})(\gamma^{v_0 \dots v_3 x_0 x_1})({\gamma^{w_0 \dots w_3}}_{x_0 x_1}) \;\; {G_{u_0 u_1}}^{y_1 y_2} G_{u_2 u_3 y_1 y_2} G_{v_0 \dots v_3} G_{w_0 \dots w_3} \\
&\tfrac{4}{24^2 \; 6^4} \; (\gamma^{u_0 u_1})(\gamma^{u_2 u_3})(\gamma^{v_0 \dots v_5})(\gamma^{w_0 \dots w_5}) \;\; G_{u_0 u_1 v_0 v_1} G_{u_2 u_3 w_0 w_1} G_{v_2 \dots v_5} G_{w_2 \dots w_5} \\
&\tfrac{4}{24^3 \; 6^4} \; (\gamma^{u_0 u_1})(\gamma^{v_0 \dots v_5})(\gamma^{w_0 \dots w_3 y_0 y_1})({\gamma^{x_0 \dots x_3}}_{y_0 y_1}) \;\; G_{u_0 u_1 v_0 v_1} G_{v_2 \dots v_5} G_{w_0 \dots w_3} G_{x_0 \dots x_3} \\
&\tfrac{1}{24^4 \; 6^4} \; (\gamma^{u_0 \dots u_3 y_0 y_1})({\gamma^{v_0 \dots v_3}}_{y_0 y_1})(\gamma^{w_0 \dots w_3 z_0 z_1})({\gamma_{x_0 \dots x_3}}_{z_0 z_1}) \;\; G_{u_0 \dots u_3} G_{v_0 \dots v_3} G_{w_0 \dots w_3} G_{x_0 \dots x_3}~,
\end{align*}
while the term $\mathrm{tr}R^4$ reads,
\begin{align*}
&\tfrac{1}{6^4} \; (\gamma^{u_0 u_1})(\gamma^{u_2 u_3})(\gamma^{u_4 u_5})(\gamma^{u_6 u_7}) \;\; {G_{u_0 u_1}}^{y_0 y_1} G_{u_2 u_3 y_0 z_0} G_{u_4 u_5 y_1 z_1} {G_{u_6 u_7}}^{z_0 z_1} \\
&\tfrac{4}{24 \cdot 6^4} \; (\gamma^{u_0 u_1})(\gamma^{u_2 u_3})(\gamma^{u_4 u_5})(\gamma^{v_0 \dots v_5}) \;\; {G_{u_0 u_1}}^{y_0 y_1} G_{u_2 u_3 v_0 y_0} G_{u_4 u_5 v_1 y_1} G_{v_2 \dots v_5} \\
&\tfrac{2}{24^2 \; 6^4} \; (\gamma^{u_0 u_1})(\gamma^{u_2 u_3})(\gamma^{v_0 \dots v_5})(\gamma^{w_0 \dots w_5}) \;\; G_{u_0 u_1 v_0 w_0} G_{u_2 u_3 v_1 w_1} G_{v_2 \dots v_5} G_{w_2 \dots w_5} \\
&\tfrac{4}{24^2 \; 6^4} \; (\gamma^{u_0 u_1})(\gamma^{u_2 u_3})(\gamma^{v_0 \dots v_4 x_0})({\gamma^{w_0 \dots w_4}}_{x_0}) \;\; {G_{u_0 u_1 v_0}}^{y_0} G_{u_2 u_3 w_0 y_0} G_{v_1 \dots v_4} G_{w_1 \dots w_4} \\
&\tfrac{4}{24^3 \; 6^4} \; (\gamma^{u_0 u_1})(\gamma^{v_0 \dots v_4 y_0})(\gamma^{w_0 \dots w_4 y_1})({\gamma^{x_0 \dots x_3}}_{y_0 y_1}) \;\; G_{u_0 u_1 v_0 w_0} G_{x_0 \dots x_3} G_{v_1 \dots v_4} G_{w_1 \dots w_4} \\
&\tfrac{1}{24^4 \; 6^4} \; ({\gamma^{u_0 \dots u_3}}_{y_0 y_1})({\gamma^{v_0 \dots v_3 y_0}}_{z_0})({\gamma^{w_0 \dots w_3 y_1}}_{z_1})({\gamma_{x_0 \dots x_3}}^{z_0 z_1}) \;\; G_{u_0 \dots u_3} G_{v_0 \dots v_3} G_{w_0 \dots w_3} G_{x_0 \dots x_3}.
\end{align*}
Suppose now that the purely femionic component of $X_8$ can be cast in the $\tau$-exact form of eq.~(\ref{te}). 
The right-hand side of eq.~(\ref{eq_Kd4L3}) would then take the form,
\begin{align*}
(\gamma_{a_1 a_2})_{\alpha_1 \alpha_2} \; X^{(8)}_{\alpha_3 \dots \alpha_{10}} &= (\gamma_{a_1 a_2})_{\alpha_1 \alpha_2} \; (\gamma^{f})_{\alpha_3 \alpha_4} \; G_{f \alpha_5 \dots \alpha_{10}} \\
&= (\gamma^{f})_{\alpha_1 \alpha_2} \Big( 3 \; (\gamma_{[a_1 a_2|})_{\alpha_3 \alpha_4} \; G_{|f] \alpha_5 \dots \alpha_{10}} - 2\;  (\gamma_{f a_1})_{\alpha_3 \alpha_4} \; G_{a_2 \alpha_5 \dots \alpha_{10}} \Big) \\
&= (\gamma^{f})_{\alpha_1 \alpha_2} \; \Big( 3 \; (\gamma_{[a_1 a_2|})_{\alpha_3 \alpha_4} \; G_{|f] \alpha_5 \dots \alpha_{10}} \Big)~,
\end{align*}
which yields, 
\eq{K_{f a_1 a_2 \alpha_3 \dots \alpha_{10}} = \; 3 \; (\gamma_{[a_1 a_2|})_{\alpha_3 \alpha_4} \; G_{|f] \alpha_5 \dots \alpha_{10}}~.\nonumber}
In the following we will 
examine whether $X_{0,8}$ can be $\tau$-exact. 
Since (\ref{eq_Kd4L3}) contains many different types of terms, it is useful to reduce this expression by simplifying every pair of $\gamma$-matrices whose bosonic indices contain contractions, using the decompositions in appendix \ref{app:gamma}.  When applied to $(\mathrm{tr}R^2)^2$, this method will give three terms of the form $\gamma^{(2)} \gamma^{(2)} \gamma^{(2)} \gamma^{(2)}$, $\gamma^{(2)} \gamma^{(2)} \gamma^{(2)} \gamma^{(6)}$ and $\gamma^{(2)} \gamma^{(2)} \gamma^{(6)} \gamma^{(6)}$, together 
with several manifestly $\tau$-exact terms. Applied to $\mathrm{tr} R^4$, this method will give several of terms of the form previously encountered, plus some new terms of the form 
$\gamma^{(2)} \gamma^{(2)} \gamma^{(5)} \gamma^{(5)}$, which are equivalent to $\gamma^{(2)} \gamma^{(2)} \gamma^{(6)} \gamma^{(6)}$ by Hodge duality. In order to compare 
$(\mathrm{tr}R^2)^2$ with $\mathrm{tr}R^4$, all the $\gamma^{(6)} \gamma^{(6)}$ terms must be converted into the form $\gamma^{(5)} \gamma^{(5)}$. This creates new $\gamma$-matrices with contracted  bosonic indices, which are simplified as before using appendix \ref{app:gamma}.   At the end of this process all the terms have the form 
$\gamma^{(2)} \gamma^{(2)} \gamma^{(2)} \gamma^{(2)}$, $\gamma^{(2)} \gamma^{(2)} \gamma^{(2)} \gamma^{(6)}$ or $\gamma^{(2)} \gamma^{(2)} \gamma^{(5)} \gamma^{(5)}$ contracted with $G^4$ (without any contractions among $\gamma$-matrices),
\begin{align}
(\gamma^{a_1 a_2})(\gamma^{a_3 a_4})(\gamma^{a_5 a_6})(\gamma^{a_7 a_8}) \; & 
G^4_{a_1 a_2;a_3 a_4;a_5 a_6;a_7 a_8}\label{eq_lastexp1} \\
(\gamma^{a_1 a_2})(\gamma^{a_3 a_4})(\gamma^{a_5 a_6})(\gamma^{a_7 \dots a_{12}}) \; & G^4_{a_1 a_2;a_3 a_4;a_5 a_6;a_7 \ldots a_{12}} \label{eq_lastexp2} \\
(\gamma^{a_1 a_2})(\gamma^{a_3 a_4})(\gamma^{a_5 \dots a_9})(\gamma^{a_{10}\ldots a_{14}}) \; &  G^4_{a_1 a_2;a_3 a_4;a_5 \dots a_9;a_{10} \ldots a_{14}}
~,
\label{eq_lastexp3}
\end{align}
up to manifestly $\tau$-exact terms which we do not need to write out explicitly. In the above,  
 $G^4_{a_1 a_2;\ldots;a_7 a_8}$, $G^4_{a_1 a_2;\ldots;a_7 \ldots a_{12}}$, $G^4_{a_1 a_2;\ldots;a_{10} \ldots a_{14}}$, denote 
certain sums of $G^4$ terms with 8,4,2 indices contracted respectively. More explicitly,
\eq{\spl{\label{c9}
G^4_{a_1 a_2;a_3 a_4;a_5 a_6;a_7 a_8}&=\frac{7}{2^73^3} ~\!G_{a_1}{}^{efg} G_{a_2a_7a_8e}G_{a_3a_5a_6}{}^hG_{a_4fgh}+\cdots\\
G^4_{a_1 a_2;a_3 a_4;a_5 a_6;a_7 \ldots a_{12}} &= \frac{25}{2^93^4} ~\!  G_{a_1a_7}{}^{fg}G_{a_2a_{11}a_{12}f} G_{a_3a_4a_9a_{10}}G_{a_5a_6a_8g}+\cdots   \\
G^4_{a_1 a_2;a_3 a_4;a_5 \dots a_9;a_{10} \ldots a_{14}}&=
\frac{1}{2^{11}3^6} ~\!  G_{a_1a_2a_{10}}{}^{f}G_{a_3a_5a_{11}f} G_{a_4a_{12}a_{13}a_{14}}G_{a_6a_7a_8a_9}+\cdots 
~,}}
where the ellipses stand for more than a hundred terms of this form. 
No obvious cancellations appear between these three types of terms at this point.

Let us further analyse how $X_{0,8}$ is decomposed into irreducible components. First, the product of four $\gamma$-matrices contains a symmetric product of eight  spinor indices which can be decomposed as follows in irreps of $B_5$,
\begin{align*}
(00001)^{\otimes_{S} 8} = \underbrace{1 (00000) \oplus \dots \oplus 1 (40000)}_{45 \text{ terms with multiplicity 1}} \oplus \underbrace{2(00004) \oplus 2(10002) \oplus 2(01002)}_{3 \text{ terms with multiplicity 2}}
~.
\end{align*}
Each irrep on the right-hand side above corresponds to a $\gamma$-structure which can be thought of as a Clebsch-Gordan coefficient: the $\gamma$-structure corresponding to 
$(00000)$ can be thought of as a Clebsch-Gordan coefficient from the scalar to $(00001)^{\otimes_{S} 8}$, etc.

Next, the product of four four-forms $G$ can be decomposed as follows in irreps of $B_5$,
\begin{align*}
(00010)^{\otimes_{S} 4} &= \underbrace{4 (00000) \oplus \dots \oplus 6(00004) \oplus \dots \oplus 3(40000)}_{95 \text{ terms, various multiplicites}}~,
\end{align*}
and all 95 terms except $(00006)$, $(00008)$, $(01006)$, and $(10006)$ can be found in $(00001)^{\otimes_{S} 8}$. This analysis implies that the contraction of four $\gamma$-matrices with four fourforms $G$ can be decomposed into $51$ $\gamma$-structures, each contracted with (multiple) $G^4$ terms corresponding to the same irrep of $B_5$. 

For example, the term $(00000)$ in the decomposition of $(00001)^{\otimes_{S} 8}$ gives rise to a single 
$\gamma$-structure contracted with the four possible $G^4$ terms giving rise to a scalar. Explicitly we have,
\begin{align*}
(\gamma^{e_1}) (\gamma_{e_1}) & (\gamma^{e_2}) (\gamma_{e_2}) \; \bigg(
\alpha_1 \; G_{a_1 \dots a_4} \; G^{a_1 \dots a_4} \; G_{b_1 \dots b_4} \; G^{b_1 \dots b_4} +
\alpha_2 \; {G_{a_1 a_2}}^{b_1 b_2} \; G_{b_1 b_2 c_1 c_2} \; {G^{c_1 c_2}}_{d_1 d_2} \; G^{d_1 d_2 a_1 a_2} \\
& + \alpha_3 \; {G_{a_1 b_1}}^{c_1 c_2} \; G_{c_1 c_2 d_1 f_1} \; {G^{a_1 d_1}}_{g_1 g_2} \; G^{g_1 g_2 b_1 f_1} +
\alpha_4 \; {G_{a_1}}^{b_1 \dots b_3} \; G_{b_1 \dots b_3 c_1} \; {G^{a_1}}_{d_1 \dots d_3} \; G^{c_1 d_1 \dots d_3} \bigg)
~,
\end{align*}
for some constants $\alpha_1,\dots,\alpha_4$. 
Similarly,  the $(00004)$ gives rise to the following term,
\begin{align*}
&\bigg(\beta_1 \;  (\gamma^{e_1}) (\gamma_{e_1}) (\gamma^{a_1 \dots a_6}) (\gamma^{b_1 \dots b_6}) + \beta_2 \; (\gamma^{[a_1}) (\gamma^{a_2 \dots a_6]}) (\gamma^{[b_1})(\gamma^{b_2 \dots b_6]}) \bigg) \\
&\times \bigg( \alpha_1 \; {G_{a_1 a_2 b_1}}^{e_1} \; G_{a_3 b_2 b_3 e_1} \; {G_{a_4 \dots a_6}}^{e_2} \; G_{b_4 \dots b_6 e_2} + \alpha_2 \; {G_{a_1 a_2 b_1}}^{e_1} \; {G_{a_3 a_4 b_2}}^{e_2} \; G_{a_5 b_3 b_4 e_1} \; G_{a_6 b_5 b_6 e_2} \\ & \qquad + \alpha_3 \; G_{a_1 \dots a_4} \; {G_{a_5 b_3 b_4}}^{e_1} \; {G_{a_6 b_5 b_6}}^{e_2} \; G_{b_1 b_2 e_1 e_2} + \alpha_4 \; {G_{a_1 b_1 b_2}}^{e_1} \; G_{a_2 \dots a_5} \; G_{a_6 b_6 e_1 e_2} \; G_{b_3 \dots b_5 e_1} \\ &\qquad + \alpha_5 \; G_{a_1 \dots a_4} \; {G_{a_5 b_1}}^{e_1 e_2} \; G_{a_6 b_2 e_1 e_2} \; G_{b_3 \dots b_6} + \alpha_6 \; G_{a_1 \dots a_3 b_3} \; G_{a_4 \dots a_6 b_2} \; {G_{b_1 b_2}}^{e_1 e_2} \; G_{b_5 b_6 e_1 e_2} \bigg)
~,
\end{align*}
for some constants $\beta_1,\beta_2,\alpha_1,\dots,\alpha_6$. 
The 51 $\gamma$-structures involved in the decomposition of $X^{(8)}$ can all be found explicitly, and only three  of them are not $\tau$-exact: $(04000)$, $(03002)$, and $(02004)$. 
In other words, except for the structures corresponding to these three irreps all other  $\gamma$-structures appearing in $X_{0,8}$ involve at least one contraction with a $\gamma^{(1)}$.

Going back to (\ref{eq_lastexp1}): 
the $G^4_{a_1a_2;\ldots;a_7a_8}$ term, by virtue of its contraction with the four $\gamma$-matrices, transforms in the symmetrized product 
of four Young  diagrams $\scalebox{0.4}{\yng(1,1)}~\!$, cf.~appendix \ref{app:Young}. Decomposing in irreducible representations of $S_8$,
\begin{align}
\label{dc1}
\scalebox{0.7}{\yng(1,1)}^{\; \otimes_{S} 4}
\ = \
\scalebox{0.7}{\yng(2,2,2,2)} \ \oplus \ \scalebox{0.7}{\yng(1,1,1,1,1,1,1,1)} \ \oplus \ \scalebox{0.7}{\yng(2,2,1,1,1,1)} \ \oplus \scalebox{0.7}{\yng(3,3,1,1)} \ \oplus \ \underbrace{\scalebox{0.7}{\yng(4,4)}}_{YT1}
\qquad \qquad (5 \text{ terms})
~.
\end{align}
At the same time $G^4_{a_1a_2;\ldots;a_7a_8}$ admits a decomposition into modules of $B_5\times S_8$, $\sum_{R} V_R \times R$,  
where $V_R$ is the plethysm of the module $V=(10000)$ of $B_5$ with respect to the Young diagram $R$ of $S_8$.  
Moreover only the plethysms corresponding to the right-hand side of (\ref{dc1}) will appear in the decomposition of 
$G^4_{a_1a_2;\ldots;a_7a_8}$ under $B_5\times S_8$. On the other hand we can compute the module $V_R$ corresponding to each $R$ 
on the right-hand side of (\ref{dc1}), using \cite{Cohen:1998}, with the result that only the plethysm corresponding to {\small $YT1$} contains $(04000)$, while neither $(02004)$ nor $(03002)$ is 
contained in any of the plethysms corresponding to the Young diagrams on the right-hand side of (\ref{dc1}).

The $G^4_{a_1a_2;\ldots;a_7\ldots a_{12}}$ term of  (\ref{eq_lastexp2}) admits the following decomposition in 
irreps of $S_{12}$,
\begin{align}\label{dc2}
\scalebox{0.7}{\yng(1,1)}^{\; \otimes_{S} 3} \ \otimes \ \scalebox{0.7}{\yng(1,1,1,1,1,1)}
\ = \
\underbrace{\scalebox{0.7}{\yng(4,4,1,1,1,1)}}_{YT2} \oplus \dots
\qquad \qquad (16 \text{ terms})~.
\end{align}
Only the plethysms corresponding to the Young diagrams on the right-hand side of (\ref{dc2}) will appear in the decomposition of 
$G^4_{a_1a_2;\ldots;a_7\ldots a_{12}}$ under $B_5\times S_{12}$. On the other hand it can be shown 
that only the plethysm corresponding to {\small $YT2$} contains $(03002)$, while neither $(04000)$ nor $(02004)$ is 
contained in any of the plethysms corresponding to the Young diagrams on the right-hand side of (\ref{dc2}).

Finally, the $G^4_{a_1a_2;\ldots;a_{10}\ldots a_{14}}$ term of  (\ref{eq_lastexp3}) admits the following decomposition in 
irreps of $S_{14}$,
\begin{align}\label{dc3}
\scalebox{0.7}{\yng(1,1)}^{\; \otimes_{S} 2} \ \otimes \ \scalebox{0.7}{\yng(1,1,1,1,1)}^{\; \otimes_{S} 2}
\ = \
\underbrace{\scalebox{0.7}{\yng(4,4,2,2,2)}}_{YT3} \ \oplus \ \underbrace{\scalebox{0.7}{\yng(4,4,2,1,1,1,1)}}_{YT4} \ \oplus \ \dots
\qquad \qquad (23 \text{ terms})~.
\end{align}
Moreover only the plethysms corresponding to the Young diagrams on the right-hand side of (\ref{dc3}) will appear in the decomposition of 
$G^4_{a_1a_2;\ldots;a_{10}\ldots a_{14}}$ under $B_5\times S_{14}$. On the other hand it can be shown 
that only the plethysm corresponding to {\small $YT3$} contains $(02004)$;  
only the plethysm corresponding to {\small $YT4$} contains $(03002)$, while $(04000)$ is not  
contained in any of the plethysms corresponding to the Young diagrams on the right-hand side of (\ref{dc3}).

Using the method of appendix \ref{app:Young}, the $\gamma$-matrices in (\ref{eq_lastexp1}) and (\ref{eq_lastexp3}) can be projected respectively onto {\small $YT1$} and {\small $YT3$}. The terms (\ref{eq_lastexp1}), (\ref{eq_lastexp3}) can thus be shown to vanish. Moreover, it can be seen that the cancellations are sensitive to the relative coefficient between $(\mathrm{tr}R^2)^2$ and $\mathrm{tr}R^4$ inside $X_8$. In other words, it can be shown that $(\mathrm{tr}R^2)^2$ and $\mathrm{tr}R^4$ are not separately $\tau$-exact.


\section{Eleven-dimensional $\gamma$-matrices}\label{app:gamma}

In this section we give our conventions for the eleven-dimensional  $\gamma$-matrices, and list a number of Fierz identities used in the analysis presented in the 
main text.

Hodge duality for $\gamma$-matrices  is defined as follows,
\eq{\star\gamma^{(n)}\label{eq_hodge}
=-(-1)^{\frac12n(n-1)}\gamma^{(11-n)}
~,}
where our definition of the Hodge operator reads,
$$ (\star S)_{a_1 \dots a_k} = \frac{1}{(11-k)!} \; {\epsilon_{a_1 \dots a_k}}^{b_1 \dots b_{11-k}} \; S_{b_1 \dots b_{11-k}}~.$$
The symmetry properties of the $\gamma$-matrices are given by,
$$(\gamma^{a_1 \dots a_n})_{\alpha \beta} =  (-1)^{\frac12(n-1)(n-2)}(\gamma^{a_1 \dots a_n})_{\beta \alpha}~,$$
where $\gamma^{(0)}$ is identified with the charge conjugation matrix.

The following Fierz identities were used in the analysis. 
Antisymmetrisation over the $a_i$ and $b_j$ indices is always understood, as well as symmetrization over all fermionic indices of the 
$\gamma$-matrices (which are suppressed here to avoid cluttering the notation),

\vspace{-0.5cm}
\begin{minipage}[c]{0.48\linewidth}
\begin{align*}
&(\gamma^{a_1 \dots a_5 e_1})(\gamma_{b_1 \dots b_5 e_1}) = \\
& + \; 120 \; \delta_{b_1 \dots b_5}^{a_1 \dots a_5} \; (\gamma^{e_1})(\gamma_{e_1}) \\
& + \; 1 \; (\gamma^{a_1 \dots a_5})(\gamma_{b_1 \dots b_5}) \\
& - \; 600 \; \delta_{b_1 \dots b_3}^{a_1 \dots a_3} \; (\gamma_{e_1})({\gamma^{e_1 a_4 a_5}}_{b_4 b_5}) \\
& + \; 25 \; \delta_{b_1}^{a_1} \; (\gamma_{e_1})({\gamma^{e_1 a_1 \dots a_4}}_{b_1 \dots b_4}) \\
& - \; 150 \; \frac{1}{2} \bigg( \delta_{b_1}^{a_1} \; (\gamma^{a_2 a_3}) ({\gamma^{a_4 a_5}}_{b_2 \dots b_5}) + (a \leftrightarrow b) \bigg) \\
& + \; 600 \; \delta_{b_1 \dots b_3}^{a_1 \dots a_3} \; (\gamma^{a_4 a_5})(\gamma_{b_4 b_5})
\end{align*}
\end{minipage}
\hfill
\begin{minipage}[c]{0.48\linewidth}
\begin{align*}
&(\gamma^{a_1 \dots a_4 e_1 e_2})(\gamma_{b_1 \dots b_4 e_1 e_2}) = \\
& - \; 12 \; \frac{1}{2} \bigg( (\gamma^{a_1 a_2}) ({\gamma^{a_3 a_4}}_{b_1 \dots b_4}) + (a \leftrightarrow b) \bigg) \\
& + \; 288 \; \delta_{b_1 b_2}^{a_1 a_2} \; (\gamma^{a_3 a_4}) (\gamma_{b_3 b_4}) \\
& - \; 96 \; \frac{1}{2} \bigg( \delta_{b_1}^{a_1} (\gamma^{a_2})({\gamma^{a_3 a_4}}_{b_2 \dots b_4})+ (a \leftrightarrow b) \bigg) \\
& + \; 192 \; \delta_{b_1 \dots b_3}^{a_1 \dots a_3} \; (\gamma^{a_4})(\gamma_{b_4}) \\
& + \; 2 \; (\gamma_{e_1})({\gamma^{e_1 a_1 \dots a_4}}_{b_1 \dots b_4}) \\
& - \; 144 \; \delta_{b_1 b_2}^{a_1 a_2} \; (\gamma_{e_1})({\gamma^{e_1 a_3 a_4}}_{b_3 b_4}) \\
& + \; 48 \; \delta_{b_1 \dots b_4}^{a_1 \dots a_4} \; (\gamma^{e_1})(\gamma_{e_1})
\end{align*}
\end{minipage}
\\
\begin{minipage}[c]{0.48\linewidth}
\begin{align*}
&(\gamma^{a_1 \dots a_3 e_1 \dots e_3})(\gamma_{b_1 \dots b_3 e_1 \dots e_3}) = \\
& + \; 36 \; \delta_{b_1 \dots b_3}^{a_1 \dots a_3} \; (\gamma_{e_1}) (\gamma^{e_1}) \\
& - \; 108 \; \delta_{b_1}^{a_1} \; (\gamma_{e_1}) ({\gamma^{e_1 a_2 a_3}}_{b_2 b_3}) \\
& + \; 216 \; \delta_{b_1}^{a_1} \; (\gamma^{a_2 a_3}) (\gamma_{b_2 b_3}) \\
& - \; 36 \; \frac{1}{2} \bigg( (\gamma^{a_1}) ({\gamma^{a_2 a_3}}_{b_1 \dots b_3}) + (a \leftrightarrow b) \bigg) \\
& + \; 324 \; \delta_{b_1 b_2}^{a_1 a_2} \; (\gamma^{a_3}) (\gamma_{b_3})
\end{align*}
\end{minipage}
\hfill
\begin{minipage}[c]{0.48\linewidth}
\begin{align*}
&(\gamma^{a_1 a_2 e_1 \dots e_4})({\gamma_{b_1 b_2}}_{e_1 \dots e_4}) = \\
& + \; 48 \; \delta_{b_1 b_2}^{a_1 a_2} \; (\gamma_{e_1}) (\gamma^{e_1}) \\
& - \; 96 \; (\gamma_{e_1}) ({\gamma^{e_1 a_1 a_2}}_{b_1 b_2}) \\
& + \; 168 \; (\gamma^{a_1 a_2}) (\gamma_{b_1 b_2}) \\
& + \; 672 \; \delta_{b_1}^{a_1} \; (\gamma^{a_1}) (\gamma^{b_1})
\end{align*}
\end{minipage}
\\
\begin{minipage}[c]{0.48\linewidth}
\begin{align*}
(\gamma^{a_1 e_1 \dots e_5})&(\gamma_{b_1 e_1 \dots e_5}) = \\
& + \; 240 \; \delta_{b_1}^{a_1} \; (\gamma_{e_1}) (\gamma^{e_1}) \\
& + \; 1680 \; (\gamma^{a_1}) (\gamma_{b_1})
\end{align*}
\end{minipage}
\hfill
\begin{minipage}[c]{0.48\linewidth}
\begin{align*}
(\gamma^{e_1 \dots e_6})&(\gamma_{e_1 \dots e_6}) = \\
& + \; 4320 \; (\gamma_{e_1}) (\gamma^{e_1})
\end{align*}
\end{minipage}
\\
\begin{minipage}[c]{0.48\linewidth}
\begin{align*}
(\gamma^{a_1 a_2 e_1 \dots e_3})&(\gamma_{b_1 b_2 e_1 \dots e_3}) = \\
& - \; 36 \; \delta_{b_1 b_2}^{a_1 a_2} \; (\gamma_{e_1}) (\gamma^{e_1}) \\
& + \; 24 \; (\gamma_{e_1}) ({\gamma^{e_1 a_1 a_2}}_{b_1 b_2}) \\
& - \; 42 \; (\gamma^{a_1 a_2}) (\gamma_{b_1 b_2}) \\
& + \; 168 \; \delta_{b_1}^{a_1} \; (\gamma^{a_2}) (\gamma_{b_2})
\end{align*}
\end{minipage}
\hfill
\begin{minipage}[c]{0.48\linewidth}
\begin{align*}
(\gamma^{a_1 e_1 \dots e_4})&({\gamma_{b_1}}_{e_1 \dots e_4}) = \\
& - \; 96 \; \delta_{b_1}^{a_1} \; (\gamma_{e_1}) (\gamma^{e_1}) \\
& + \; 336 \; (\gamma^{a_1}) (\gamma_{b_1})
\end{align*}
\begin{align*}
(\gamma^{e_1 \dots e_5})&(\gamma_{e_1 \dots e_5}) = \\
& - \; 720 \; (\gamma_{e_1}) (\gamma^{e_1})
\end{align*}
\end{minipage}
\\
\begin{minipage}[c]{0.48\linewidth}
\begin{align*}
&(\gamma^{a_1 \dots a_4 e_1})({\gamma_{b_1 \dots b_4}}_{e_1}) = \\
& + \; 6 \; \frac{1}{2} \bigg( (\gamma^{a_1 a_2}) ({\gamma^{a_3 a_4}}_{b_1 \dots b_4}) + (a \leftrightarrow b) \bigg) \\
& - \; 72 \; \delta_{b_1 b_2}^{a_1 a_2} \; (\gamma^{a_3 a_4}) (\gamma_{b_3 b_4}) \\
& - \; 48 \; \frac{1}{2} \bigg( \delta_{b_1}^{a_1} \; (\gamma^{a_2})({\gamma^{a_3 a_4}}_{b_2 \dots b_4})+ (a \leftrightarrow b) \bigg) \\
& + \; 96 \; \delta_{b_1 \dots b_3}^{a_1 \dots a_3} \; (\gamma^{a_4})(\gamma_{b_4}) \\
& - \; 1 \; (\gamma_{e_1})({\gamma^{e_1 a_1 \dots a_4}}_{b_1 \dots b_4}) \\
& + \; 72 \; \delta_{b_1 b_2}^{a_1 a_2} \; (\gamma_{e_1})({\gamma^{e_1 a_3 a_4}}_{b_3 b_4}) \\
& - \; 24 \; \delta_{b_1 \dots b_4}^{a_1 \dots a_4} \; (\gamma^{e_1})(\gamma_{e_1})
\end{align*}
\end{minipage}
\hfill
\begin{minipage}[c]{0.48\linewidth}
\begin{align*}
&(\gamma^{a_1 \dots a_3 e_1 e_2})(\gamma_{b_1 \dots b_3 e_1 e_2}) = \\
& - \; 24 \; \delta_{b_1 \dots b_3}^{a_1 \dots a_3} \; (\gamma_{e_1}) (\gamma^{e_1}) \\
& + \; 36 \; \delta_{b_1}^{a_1} \; (\gamma_{e_1}) ({\gamma^{e_1 a_2 a_3}}_{b_2 b_3}) \\
& - \; 54 \; \delta_{b_1}^{a_1} \; (\gamma^{a_2 a_3}) (\gamma_{b_2 b_3}) \\
& - \; 12 \; \frac{1}{2} \bigg( (\gamma^{a_1}) ({\gamma^{a_2 a_3}}_{b_1 \dots b_3}) + (a \leftrightarrow b) \bigg) \\
& + \; 108 \; \delta_{b_1 b_2}^{a_1 a_2} \; (\gamma^{a_3}) (\gamma_{b_3})
\end{align*}
\end{minipage}
\\
\begin{minipage}[c]{0.48\linewidth}
\begin{align*}
(\gamma^{a_1 e_1})&({\gamma_{b_1 \dots b_4}}_{e_1}) = \\
& + \; 1 \; (\gamma_{e_1})({\gamma^{e_1 a_1}}_{b_1 \dots b_4}) \\
& + \; 12 \; \delta_{b_1}^{a_1} \; (\gamma_{b_2})(\gamma_{b_3 b_4})
\end{align*}
\end{minipage}
\hfill
\begin{minipage}[c]{0.48\linewidth}
\begin{align*}
(\gamma^{a_1 e_1})(\gamma_{b_1 e_1}) =
& + \; 1 \; (\gamma^{a_1})(\gamma_{b_1}) \\
& - \; 1 \; \delta_{b_1}^{a_1} \; (\gamma_{e_1})(\gamma^{e_1})
\end{align*}
\end{minipage}
\\
\begin{minipage}[c]{0.48\linewidth}
\begin{align*}
(\gamma^{a_1 \dots a_4 e_1})&({\gamma^{b_1 \dots b_5}}_{e_1}) = \\
& - \; 60 \; \eta^{a_1 b_1} \; (\gamma^{a_2 a_3}) (\gamma^{a_4 b_2 \dots b_5}) \\
& - \; 60 \; \eta_{a_1 b_1} \; (\gamma^{a_2}) (\gamma^{a_3 a_4 b_2 \dots b_5}) \\
& - \; 720 \; \delta^{a_1 \dots a_3}_{c_1 \dots c_3} \; \eta^{c_1 b_1} \eta^{c_2 b_2} \eta^{c_3 b_3} \; (\gamma^{a_4}) (\gamma^{b_4 b_5}) \\
& + \; 240 \; \delta^{a_1 \dots a_3}_{c_1 \dots c_3} \; \eta^{c_1 b_1} \eta^{c_2 b_2} \eta^{c_3 b_3} \; (\gamma^{a_4}) (\gamma^{b_4 b_5}) \\
& + \; 140 \; \eta^{a_1 b_1} \; (\gamma^{[a_2}) (\gamma^{a_3 a_4 b_2 \dots b_5]}) \\
& - \; 120 \; \delta^{a_1 a_2}_{c_1 c2} \; \eta^{c_1 b_1} \eta^{c_2 b_2}\; (\gamma_{e_1})(\gamma^{e_1 a_3 a_4 b_3 \dots b_5})
\end{align*}
\end{minipage}
\hfill
\begin{minipage}[c]{0.48\linewidth}
\begin{align*}
(\gamma^{a_1 e_1})&({\gamma^{b_1 \dots b_5}}_{e_1}) = \\
& - \; 6 \; (\gamma^{[a_1})(\gamma^{b_1 \dots b_5]}) \\
& - \; 5 \; \eta^{a_1 b_1} \; (\gamma_{e_1})(\gamma^{e_1 b_2 \dots b_5}) \\
& + \; 1 \; (\gamma^{a_1})(\gamma^{b_1 \dots b_5})
\end{align*}
\end{minipage}


\section{Eleven-dimensional superspace}\label{app:ss}

In this section we review the properties of on-shell eleven-dimensional superspace  at lowest order in the Planck length \cite{Brink:1980}. The theory 
thus obtained is equivalent to CJS supergravity \cite{Cremmer:1978}.

The non-zero superfield components are as follows,
\begin{align}
G_{a b \alpha \beta} &= -i \; (\gamma_{ab})_{\alpha \beta} \notag \\
{T_{\alpha \beta}}^{f} &= -i \; (\gamma^{f})_{\alpha \beta} \notag \\
{T_{a \alpha}}^{\beta} &= -\frac{1}{36} \left( {(\gamma^{bcd})_{\alpha}}^{\beta} G_{abcd} + \frac{1}{8} {({\gamma_{a}}^{bcde})_{\alpha}}^{\beta} G_{bcde} \right) \notag \\
R_{\alpha \beta a b} &= \frac{i}{6} \left( (\gamma^{gh})_{\alpha \beta} \; G_{ghab} + \frac{1}{24} \; ({\gamma_{ab}}^{ghij})_{\alpha \beta} \; G_{ghij} \right) \notag \\
R_{\alpha a b c} &= \frac{i}{2} \left( (\gamma_{a} T_{bc})_{\alpha} - 2 (\gamma_{[b} T_{c]a})_{\alpha} \right)~.
\label{eq_fieldcomp}
\end{align}
The action of the spinorial derivative on the superfields reads,
\eq{\spl{
D_{\alpha} G_{abcd} &= 6 \;i\; (\gamma_{[ab|})_{\alpha \epsilon} \; {T_{|cd]}}^{\epsilon} \\
D_{\alpha} R_{abcd} &= \d_{[a|} R_{\alpha |b]cd} - {T_{ab}}^{\epsilon} R_{\epsilon \alpha cd} + 2 \; {T_{[a| \alpha}}^{\epsilon} R_{\epsilon |b] cd} \\
D_{\alpha} {T_{ab}}^{\beta} &= \frac{1}{4} \; R_{abcd} {(\gamma^{cd})_{\alpha}}^{\beta} - 2 \; D_{[a} {T_{b] \alpha}}^{\beta} - 2 \; {T_{[a| \alpha}}^{\epsilon} {T_{[b] \epsilon}}^{\beta}~.
\label{eq_spinder}
}}
The equations of motion for the field-strengths $G$,$R$ and $T$ are given by,
\begin{align}
D^{f} G_{f a_1 a_2 a_3} &= -\frac{1}{1152} \; \epsilon_{a_1 a_2 a_3 b_1 \dots b_4 c_1 \dots c_4} G^{b_1 \dots b_4} G^{c_1 \dots c_4} \notag \\
(\gamma^{a})_{\alpha \epsilon} \; {T_{ab}}^{\epsilon} &= 0 \notag \\
R_{ab} - \frac{1}{2} \eta_{ab} R &= \frac{1}{12} \left( G_{afgh} {G_{b}}^{fgh} - \frac{1}{8} \eta_{ab} G_{fghi} G^{fghi} \right)~.
\label{eq_eom}
\end{align}


\section{Tensor representation of a Young diagram}\label{app:Young}

A Young diagram with $n$ boxes, see \cite{Fulling:1992vm} for a review, represents an irreducible representation of the symmetric group $S_{n}$. It is possible to give  explicit expressions for Young diagrams in the form of tensors. 
The method is more easily understood using a specific example. Consider a tensor $T_{a_1 a_2 a_3 a_4}$ without any 
a priori symmetry properties, and  let us construct its projection onto  \scalebox{0.6}{$\yng(3,1)$}. Several symmetry operations will have to be applied on the tensor, but the Young diagram does not state which indices correspond to its different boxes. First one must determine all the {\em standard tableaux}, i.e. all the Young diagrams with numbered boxes, with increasing numbers in all rows and columns. Different Young tableaux corresponding to the same Young diagram give equivalent but distinct representations of the symmetric group. The diagram \scalebox{0.6}{$\yng(3,1)$} has three standard tableaux, \scalebox{0.6}{$\young(123,4)$}, \scalebox{0.6}{$\young(134,2)$}, and \scalebox{0.6}{$\young(124,3)$}, to which correspond three tensors, $T^{(1)}$, $T^{(2)}$ and $T^{(3)}$ respectively.

To obtain the tensor corresponding to a given standard tableau, one must  first symmetrize over the indices indicated in each row, and then antisymmetrize over the indices indicated in each column. 
For example, $(T^{(1)})_{a_1 a_2 a_3 a_4}$ will be obtained by first symmetrizing  over the indices $a_1$, $a_2$ and $a_3$,
\begin{align*}
\big(T_{a_1 a_2 a_3 a_4}+T_{a_1 a_3 a_2 a_4}+T_{a_2 a_1 a_3 a_4}+T_{a_2 a_3 a_1 a_4}+T_{a_3 a_1 a_2 a_4}+T_{a_3 a_2 a_1 a_4} \big)~,
\end{align*}
and then antisymmetrizing over $a_1$ and $a_4$,
\begin{align*}
&(\Pi^{(1)} T)_{a_1 a_2 a_3 a_4} = (T^{(1)})_{a_1 a_2 a_3 a_4} =\\
&\qquad \qquad  \frac18\big( T_{a_1 a_2 a_3 a_4}+T_{a_1 a_3 a_2 a_4}+T_{a_2a_1 a_3 a_4}+T_{a_2 a_3 a_1 a_4}+T_{a_2 a_3 a_4 a_1}+T_{a_2 a_4 a_3 a_1}+\\
&\qquad \qquad \quad \; T_{a_3 a_1 a_2 a_4}+T_{a_3 a_2 a_1 a_4}+T_{a_3 a_2 a_4 a_1}+T_{a_3 a_4 a_2 a_1}+T_{a_4 a_2 a_3 a_1}+T_{a_4 a_3 a_2 a_1} \big)~.
\end{align*}
The overall normalization above can be straightforwardly determined by imposing $\Pi^{(1)}\Pi^{(1)} T = \Pi^{(1)} T$, where $\Pi^{(1)} T=T^{(1)}$ is the projection of the tensor $T$ onto the Young tableau \scalebox{0.6}{$\young(123,4)$}.

For example the tensors $T^{(1)}$ and $T^{(2)}$, associated with \scalebox{0.6}{$\young(123,4)$} and \scalebox{0.6}{$\young(134,2)$} respectively, obey the following properties,
\begin{align*}
&(T^{(1)})_{[a b| c |d]} = 0 && (T^{(2)})_{[a b c] d} = 0 \\
&(T^{(1)})_{[a| b c |d]} = (T^{(1)})_{a b c d} && (T^{(2)})_{[a b] c d} = (T^{(2)})_{a b c d} \\
&(T^{(1)})_{a (b c) d} = (T^{(1)})_{a b c d} && (T^{(2)})_{a b(c d)} = (T^{(2)})_{a b c d}~.
\end{align*}
More generally, each $T^{(i)}$ has exactly  three independent orderings of indices, which can be taken to be $T^{(1)}_{a_1 a_4 a_3 a_2}$, $T^{(1)}_{a_2 a_1 a_4 a_3}$ and 
$T^{(1)}_{a_2 a_3 a_1 a_4}$. Any symmetry operation on the indices of $T^{(i)}$ can be expressed as a linear combination of these three orderings, e.g.,
\begin{align*}
T^{(1)}_{a_1 (a_2 a_3 a_4)} &= T^{(1)}_{a_1 a_4 a_3 a_2} + \frac{1}{3} \; T^{(1)}_{a_2 a_1 a_4 a_3} + \frac{1}{3} \; T^{(1)}_{a_2 a_3 a_1 a_4} \\
T^{(1)}_{[a_1 a_2] a_3 a_4} &=  \frac{1}{2} \; T^{(1)}_{a_1 a_4 a_3 a_2} + \frac{1}{2} \; T^{(1)}_{a_2 a_1 a_4 a_3} + 0 \; T^{(1)}_{a_2 a_3 a_1 a_4}~.
\end{align*}
A tensor $T$ projected onto a non-standard tableau can be expressed as a linear combination of the three standard ones. For example it is straightforward (but tedious) to check that the projection onto the non-standard tableau  \scalebox{0.6}{$\young(342,1)$} can be decomposed as,
\eq{\spl{\label{eca}
(\Pi^{(4)} T)_{a_1 a_2 a_3 a_4} &= T^{(1)}_{a_1 a_4 a_3 a_2} + T^{(1)}_{a_2 a_1 a_4 a_3} + 0 \; T^{(1)}_{a_2 a_3 a_1 a_4} \\
&+ 0 \; T^{(2)}_{a_1 a_4 a_3 a_2} - T^{(2)}_{a_2 a_1 a_4 a_3} + T^{(2)}_{a_2 a_3 a_1 a_4} \\
&+ T^{(3)}_{a_2 a_1 a_3 a_4} + 0 \; T^{(3)}_{a_2 a_1 a_4 a_3} - T^{(3)}_{a_2 a_3 a_1 a_4}~.
}}
Every other tableau (corresponding to the same Young diagram  \scalebox{0.6}{$\yng(3,1)$}) and any symmetry operation on the indices can be expressed as a linear combination of those nine elements. The automatization of general decompositions onto Young tablaux, such as the one above, has been implemented in the computer program \cite{Bertrand}.

More generally a tensor $T_{a_1 a_2 a_3 a_4}$  without any 
a priori symmetry properties can be decomposed into ten Young tableaux,
\eq{\spl{\label{geca}
\underbrace{\scalebox{0.6}{\yng(1)}^{\; \otimes 4}}_{T} = \;   \underbrace{\scalebox{0.6}{\yng(4)}}_{T^{S}}   \; \oplus \;\; 
\underbrace{3 \;\; \scalebox{0.6}{\yng(3,1)}}_{T^{(1,2,3)}} 
\; \oplus     \;\; \underbrace{2 \;\; \scalebox{0.6}{\yng(2,2)}}_{T^{\prime(1,2)}} \; \oplus \;\; 
\underbrace{3 \;\;\scalebox{0.6}{\yng(2,1,1)}}_{T^{\prime\prime(1,2,3)}} 
\; \oplus \; 
\underbrace{\scalebox{0.6}{\yng(1,1,1,1)}}_{T^{A}}
~,
}}
where $T^{(1)}$, $T^{(2)}$ and $T^{(3)}$ are the Young tableaux appearing on the right-hand side of (\ref{eca}) above,  and correspond to the term $3$ \scalebox{0.6}{$\yng(3,1)$}. The 
remaining Young tableaux in the decomposition can be explicitly constructed using the same method. 

Consider now  a tensor $T$ with a symmetry structure given by, e.g., \scalebox{0.6}{$\yng(2,1)$} $\otimes$ \scalebox{0.6}{$\yng(1)$}. 
The previous decomposition of $\scalebox{0.6}{\yng(1)}^{\; \otimes 4}$ can also be used to decompose $T$ into its irreducible components. 
Indeed, a tensor with structure \scalebox{0.6}{$\yng(2,1)$} $\otimes$ \scalebox{0.6}{$\yng(1)$} can be viewed as a particular set of symmetry operations performed on the indices of a tensor without any symmetry (i.e. with structure \scalebox{0.6}{$\yng(1)$}$^{\; \otimes 4}$). Therefore $T$ can be expressed as a linear combination of the tensors already used in the decomposition (\ref{geca}).

The following example shows the decomposition of the symmetric product of two threeforms $H$,
\begin{align*}
T_{a_1\dots a_6} :=H_{a_1 a_2 a_3} H_{a_4 a_5 a_6} \quad \longrightarrow \qquad \scalebox{0.6}{\yng(1,1,1)}^{\otimes_{S} 2} = \scalebox{0.6}{\yng(2,2,2)} \oplus \scalebox{0.6}{\yng(2,1,1,1,1)}
~.
\end{align*}
\vskip -1cm

There are five standard tableaux corresponding to each of the Young diagrams $\scalebox{0.6}{\yng(2,2,2)}$, $\scalebox{0.6}{\yng(2,1,1,1,1)}$. The tensors corresponding to these Young tableaux can be denoted by $T^{(1)},\dots T^{(5)}$ and 
$T^{\prime(1)},\dots T^{\prime(5)}$, respectively. 
In the particular example above, it can be shown  that,
\begin{align*}
H_{a_1 a_2 a_3} H_{a_4 a_5 a_6} = T^{(1)}_{a_1 a_2 a_3 a_4 a_5 a_6} + T^{\prime(1)}_{a_1 a_2 a_3 a_4 a_5 a_6} + T^{\prime(1)}_{a_1 a_2 a_3 a_4 a_6 a_5} - 
T^{\prime(1)}_{a_1 a_2 a_3 a_5 a_6 a_4}~,
\end{align*}
i.e. only the tensors $T^{(1)}$ and $T^{\prime(1)}$, corresponding to \scalebox{0.5}{$\young(14,25,36)$} and \scalebox{0.5}{$\young(16,2,3,4,5)$} respectively, 
enter the decomposition.

\vfill\break




%
%
\bibliography{refs}
\bibliographystyle{unsrt}
\end{document}